\documentclass[
reprint,onecolumn,
superscriptaddress, 12pt, 
nofootinbib,
 amsmath,amssymb,
floatfix
]{revtex4-2}

\usepackage{physics} %for real and imaginary parts
\usepackage{graphicx}% Include figure files
\usepackage{dcolumn}% Align table columns on decimal point
\usepackage{bm}% bold math
\usepackage{cancel}% to cancel terms in equations
\usepackage{amsfonts}%bold math
\usepackage{xcolor}%to color some texts
\usepackage{tcolorbox}% for boxing texts

\newcommand{\beqa}{\begin{eqnarray}}
\newcommand{\eeqa}{\end{eqnarray}}
\newcommand{\nn}{\nonumber}

\usepackage{graphicx}% Include figure files
\usepackage{dcolumn}% Align table columns on decimal point
\usepackage{bm}% bold math

\usepackage{mathtools}
\usepackage{graphicx}

\begin{document}

\title{ Dynamics of Vortex Clusters on a Torus}
\author{Aswathy K R}
\affiliation{Birla Institute of Technology and Science, Pilani, Hyderabad Campus, Telangana 500078, India}
\author{Udaya Maurya}
\affiliation{Institute for Plasma Research, Bhat, Gandhinagar, India}
\author{Surya Teja Gavva}
\affiliation {Department of Computer Science,
Rutgers University, Piscataway, NJ 08854-8019, USA}
\author{Rickmoy Samanta}
\affiliation{Birla Institute of Technology and Science, Pilani, Hyderabad Campus, Telangana 500078, India}

\begin{abstract}
Abstract: We investigate the collective dynamics of multivortex assemblies in a two dimensional  toroidal fluid film of distinct curvature and topology. The incompressible, inviscid nature of the fluid permits a Hamiltonian description of the vortices, along with a self-force of geometric origin. The Hamiltonian dynamics is constructed in terms of q-digamma functions $\Psi_q(z)$, closely related to the Schottky-Klein prime function known to arise in multiply connected domains. We show the fundamental motion of the two-vortex system and identify five classes of geodesics on the torus for the special case of a vortex dipole, along with subtle distinctions from vortices in quantum superfluids. In multivortex assemblies, we observe that a randomly initialized chiral cluster of vortices  remains geometrically confined on the torus, while undergoing an overall drift along the toroidal direction. A cluster of fast and slow vortices also show the collective toroidal drift, with the fast ones predominantly occupying the core region of the revolving cluster. Achiral clusters show unconfined dynamics and scatter all over the surface of the torus. A chiral cluster with an impurity in the form of a single vortex of opposite sign also show similar behavior as a pure chiral cluster, with occasional “jets” of dipoles leaving and re-entering the revolving cluster. The work serves as a step towards analysis of vortex clusters in models that incorporate harmonic velocities in the Hodge decomposition.
\end{abstract}
\maketitle
\section{Introduction}
A recurring theme in modern hydrodynamics is the study of microscopic interactions in an assembly of structures or point-like defects in a medium, leading to emergent collective dynamics. Often the geometrical features of the underlying medium play a crucial role in this emergent behavior.  A canonical example is that of point vortex interactions in thin two-dimensional (2D) fluid films of prescribed geometry. Although the study of point vortex dynamics in flat domains dates back to the early works of Kelvin and Helmholtz Ref.~\cite{moffatt, aref, aref1, saffman, lin1, lin2}, the study of point vortices in fluid surfaces of distinct curvature and topology is gaining increasing attention, see for example Ref~\cite{ bg, hally, kimok, kimura, khesin2024, newton, crowdymarshall,turner, voigt, boattok, boattod, moffat2014, ershkov2016, grms, sakajo2009, newtonsakajo1, newtonsakajo2, nelsonsakajo, sakajo2016, sakajo2018, sakajo2019, sakajo2023, sakajo2025}, with interesting applications to fluid membranes Ref.~\cite{rs2021, rs2022,rs2023,rs2025, lauga2025}. In this work, we investigate classical point vortex dynamics in two-dimensional fluid films of toroidal geometry Ref.~\cite{grms, sakajo2016}, which share close connections with rotors in classical fluid interfaces Ref.~\cite{lushi,yeo,sh1,sh2}. This study is thematically related to a broad range of systems recently explored in toroidal geometries, including active suspensions confined on toroidal droplets Ref.~\cite{giomi2019}, toroidal crystals Ref.~\cite{giomi2008,giomi2008pre, giomi2024}, and quantum vortices in toroidal superfluid films and porous media Ref.~\cite{feynman, emn, machta, fetter}. Vortex clusters in superfluids Ref.~\cite{abanov,vsc} have also attracted growing interest. Moreover, superfluid vortices in toroidal cold-atom traps and in other geometries Ref.~\cite{trap1, trap2, andrea1, andrea2, andrea3} are being actively investigated, with experiments planned for microgravity environments Ref.~\cite{iss1,iss2}. \\
In the present work, point vortices are modeled as point-like singularities with constant circulation in a classical, incompressible, and inviscid fluid film of toroidal shape. The mathematical formulation builds on the foundational works of Green and Marshall Ref.~\cite{grms} and Sakajo and Shimizu Ref.~\cite{sakajo2016}. Sakajo and Shimizu employed the Green’s function for toroidal surfaces, developed by Green and Marshall, to construct a Hamiltonian dynamical system for vortices on the torus, focusing primarily on two-vortex interactions and equilibrium configurations. Here, we recast the Hamiltonian dynamics in terms of well-tabulated $q$-digamma functions $\Psi_q(\zeta)$, where $q$ is determined by the torus size and $\zeta$ is a complex coordinate on the torus (to be detailed later). These functions are closely related to the Schottky-Klein prime function, which is well known to arise in multiply connected domains Ref.~\cite{crowdymarshall, crowdyskpaper, crowdybook}. This representation facilitates numerical simulations with a large number of vortices and provides a framework for studying the evolution of vortex clusters on the torus, particularly chiral clusters, which have proven to be difficult so far (Ref.~\cite{sakajo2016,sakajo2018}). \\
In the first half of the paper, we explore the two-vortex system and identify five classes of geodesics on the torus for the special case of a vortex pair of opposite sign (``\textit{vortex dipole}"). We also observe single loop and double loop trajectories for the vortex pair of same strength and circulation (``\textit{chiral vortex pair}"). We then explore multivortex assemblies.  We observe that a cluster of vortices of the same sign (``\textit{chiral cluster}") remains geometrically confined on the torus (area preserving), while undergoing an overall drift along the toroidal direction, resembling collective dynamics. A mixture of fast and slow vortices of same sign (``\textit{chiral fast-slow cluster}") also show the collective toroidal drift, with the fast ones predominantly occupying the core region and the slow ones expelled to the periphery of the revolving cluster. Vortex clusters of mixed sign but zero net circulation (``\textit{achiral or neutral cluster}") show unconfined dynamics and scatter all over the surface of the torus. A chiral cluster with an impurity in the form of a single vortex of opposite sign also show similar behavior as a pure chiral cluster, with occasional ``jets" of dipoles leaving and re-entering the revolving cluster. These results comprise the key findings of this study. \\
Let us point out that the present study differs significantly from investigations of \textit{vortex crystals} on the torus. These are special vortex configurations that lead to fixed or relative equilibria and occur only for specific arrangements of vortices and circulation strengths; see Ref.~\cite{sakajo2019} for many beautiful examples. In contrast, our objective is to explore the collective dynamics of a single vortex cluster with vortices initially placed at random locations within the cluster. Our work is more closely related to ongoing research on two-dimensional hydrodynamics of vortex fluids Ref.~\cite{vsc} and its generalization to curved spaces Ref.~\cite{turner}. It is also worth noting some key differences between the classical model considered here and vortex interactions in a toroidal superfluid film Ref.~\cite{fetter}. In the superfluid case, the single-valued nature of the condensate wavefunction introduces an additional quantum interaction. This leads to important deviations from the classical model for certain configurations, such as the diametrically opposite vortex dipole, as discussed in Ref.~\cite{fetter} and examined further in Sec. \ref{onetwovortex}. \\
On a technical note, we adopt a \textit{local} vortex interaction model based on the foundational works of Green and Marshall Ref.~\cite{grms} and Sakajo and Shimizu Ref.~\cite{sakajo2016}. The vorticity–streamfunction approach used here has limitations regarding the inclusion of harmonic velocity fields Ref.~\cite{h1,h2,h3,khesin2024}. A more complete treatment requires an enlarged phase space of dimension $2N+2g$ for an $N$-vortex system on a surface of genus $g$, which can render even the single-vortex problem non-integrable on the torus Ref.~\cite{h3}. It has been argued in Ref.~\cite{khesin2024} that for closely spaced vortices, both the descriptions agree, as the harmonic contributions are of lower order. For numerical tractability in simulating vortex clusters, we therefore restrict ourselves to the simpler model Ref.~\cite{grms, sakajo2016, sakajo2018, sakajo2019}. This work can thus be regarded as a step toward analyzing vortex clusters in models that incorporate harmonic velocities in the Hodge decomposition. \\
This article is organized as follows:  In Sec.~\ref{recap}, we introduce the background geometry of the torus and formulate the dynamical equations along with the associated conservation laws. In Sec.~\ref{onetwovortex}, we perform several consistency checks focusing on one- and two-vortex configurations, identifying five classes of geodesics for the vortex dipole. In Sec.~\ref{multivortex}, we investigate the dynamics of vortex clusters of both uniform and mixed populations, with particular emphasis on the effects of the torus’s distinct topology and curvature. We conclude in Sec.~\ref{cncl} with a discussion of future directions. Additional details and derivations are provided in Appendices \ref{app1}, \ref{app2}, \ref{app3}, and \ref{app4}.
\begin{figure}[h!]
\begin{center}
%\begin{tabular}{ccccccccc}
\includegraphics[width=0.375\linewidth]{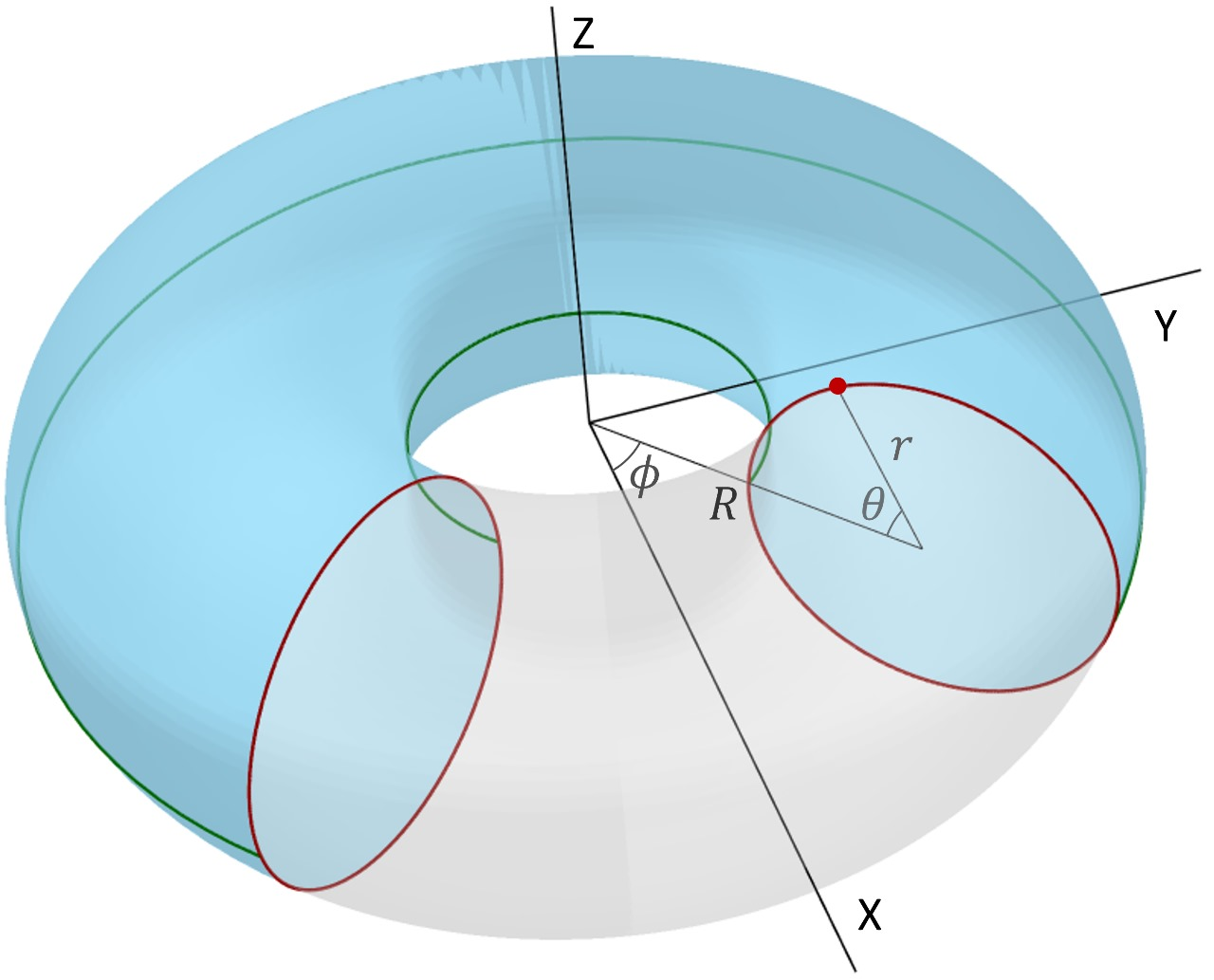}
\includegraphics[width=0.525\linewidth]{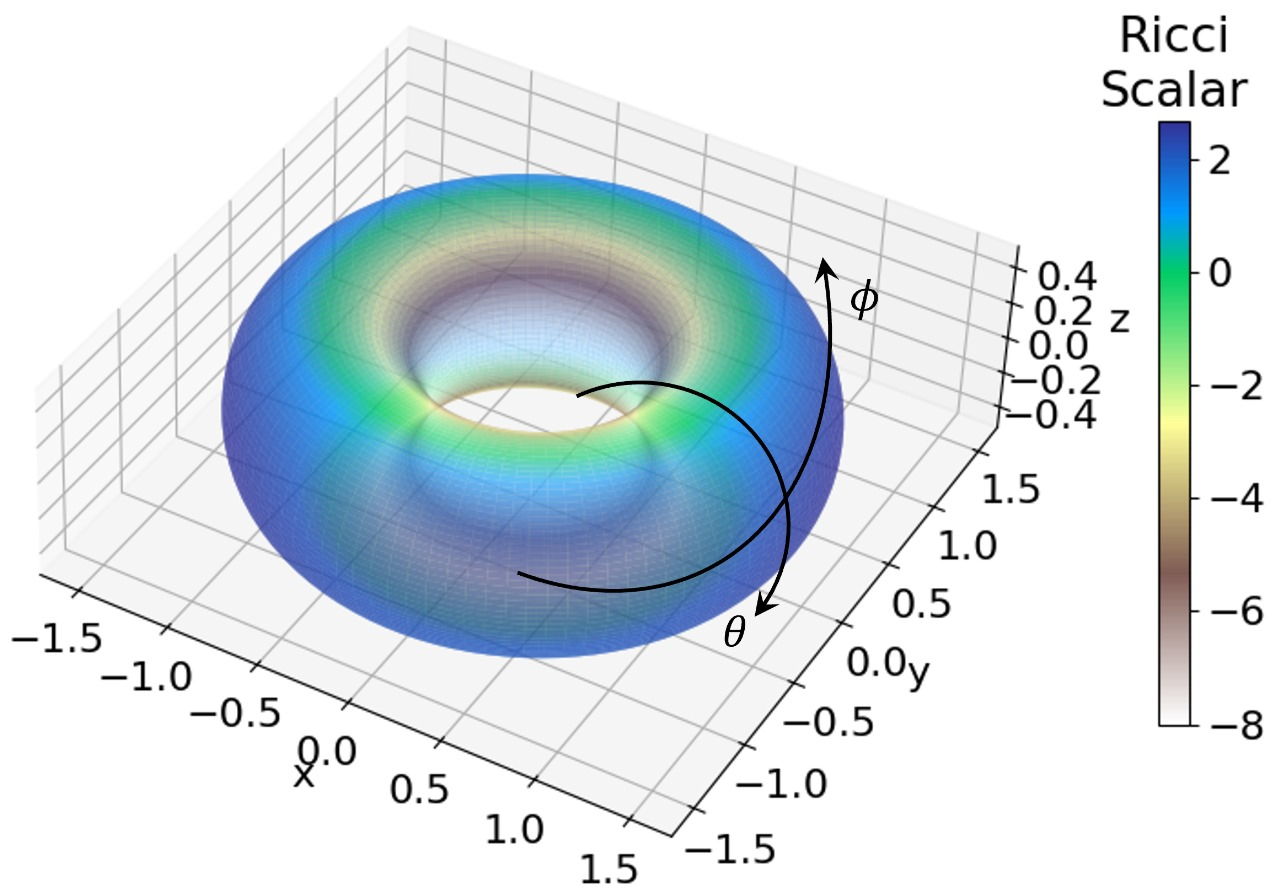}
%\end{tabular}
 \caption{Torus embedding and curvature: Left: 3D embedding of the torus as described in the main text. Right: toroidal direction (increasing $\phi$) and poloidal direction (increasing $\theta$) on the torus, together with a color map showing the local Ricci scalar curvature, whose expression is given in the main text. The parameters are chosen as $R = 1$ and $r = 0.5$, giving $\alpha = R/r = 2$ as discussed. Note that the Gaussian curvature is equal to half the Ricci scalar curvature.}
 \label{torusmap} 
\end{center}
\end{figure}
\section{Vortex model on the torus}
\label{recap}
We set up the background geometry of the torus via the standard three dimensional embedding (Ref.~\cite{grms}) described by 
\begin{eqnarray*}
    x&=&\left(R-r\cos\theta\right)\cos \phi\\
    y&=&\left(R-r\cos\theta\right)\sin \phi\\
    z&=&r\sin\theta
\end{eqnarray*}
where $R$ is the distance from the center of the tube to the center of the torus and $r$ is the radius of the tube, see Fig.~(\ref{torusmap}). Both $r$ and $R$ are kept constant throughout the paper. We will use the terminology ``toroidal motion" to indicate motion along coordinate $\phi$ and ``poloidal/meridional motion"  for motion along $\theta$ direction, as shown in Fig.~(\ref{torusmap}). The metric describing the toroidal surface is given by
\begin{equation}
    ds^2=r^2 d\theta^2+\left(R-r\cos\theta\right)^2 d\phi^2 \nn
    \label{met}
\end{equation}
and the Ricci scalar is given by
\beqa
\frac{2\cos\theta}{-rR+r^2\cos\theta}
\eeqa 
which indicates three distinct curvature regions on the torus ie. the inner equatorial negative curvature region around $\theta=0$, the flat regions around $\theta=\pm \pi/2$ and the outer positive curvature region, around $\theta = \pi$, see Fig.~(\ref{torusmap}).
\begin{comment}
\begin{figure}[h!]
\includegraphics[width=6cm]{images/ricci.png}
\label{ric}
\end{figure}
\end{comment}
Following Sakajo and Shimizu Ref.~(\cite{sakajo2016}), it will be convenient to introduce the following parameters
\beqa
\alpha=\frac{R}{r},~ A =(\alpha^2 -1)^{-\frac{1}{2}},~ c~=-\alpha-A^{-1},~ \rho =e^{-2 \pi A}.\nn
\label{param}
\eeqa
In terms of the above defined parameter $\alpha$, we see that the ratio  of the magnitudes of positive curvature on the outer equator to that of negative curvature on the inner equator of the torus is given by the ratio $\frac{\alpha-1}{\alpha+1}$. In the limit of large $\alpha$ (ie. thin tori), the ratio tends to 1, which implies that it is equally curved on the outer and inner surfaces like a cylinder (flat limit). In the opposite limit, for ``thick" tori (ie. small values of $\alpha \sim 1$), the ratio vanishes, implying high values of negative curvature on the inner surface of the torus. In this work, we set $R=1$ and $r=0.5$ such that $\alpha=2$. The vortex solution is constructed by placing a point vortex of constant circulation $\Gamma$ on the torus and find the resulting 2D velocity field on the surface of the torus. The inviscid and incompressible nature of the fluid allows the 2D vortex velocity field $\bf{u}$ to be expressed \textit{locally} in terms of a stream function $\psi$ Ref.~(\cite{ hally,boattok, boattod}) as follows
\begin{equation}
{\bf u}=\left( \frac{1}{R-r \cos \theta} \frac{\partial \psi}{\partial \phi},-\frac{1}{r} \frac{\partial \psi}{\partial \theta}\right)
\label{vsol}
\end{equation}
where $\psi$ is essentially the Green's function $G_H$  of the Laplace operator on the torus (see also Ref.~(\cite{h1,h2,h3}) for improved models which incorporate harmonic velocities as mentioned in the introduction). Schematically, the hydrodynamic Green's function $G_H$ for a point vortex of constant circulation $\Gamma$ satisfies (Ref.~\cite{grms})
  \begin{equation}
\nabla_{\mathcal{T}_{R, r}}^2 G_H= \Gamma ~ \delta-\frac{1}{4 \pi^2 r R}
\end{equation}
where $\nabla_{\mathcal{T}_{R, r}}^2$ is the Laplace  operator on the torus and we have subtracted a background of uniform vorticity such that
the circulation associated with the point vortex ($\delta$) is nullified, to be consistent with Gauss's Divergence theorem on closed surfaces  (Ref.~\cite{grms}). Following the conformal mapping techniques of Ref.~\cite{hally, boattok,boattod,grms,sakajo2016}, it is possible to find a closed-form analytic expression for the Green's function $G_H$, Ref.\cite{grms}. For this purpose, we introduce a complex coordinate $\zeta$ on the torus defined via the conformal map (Ref.~\cite{akh})
   \begin{equation}
      \zeta(\theta, \phi)  \longmapsto \mathrm{e}^{\mathrm{i} \phi} \exp \left(-\int_0^\theta \frac{\mathrm{d} u}{\alpha-\cos u}\right) \equiv \mathrm{e}^{\mathrm{i} \phi} \exp \left(r_c(\theta)\right) \label{conformalmap}
   \end{equation} 
 where
 \beqa
 r_c(\theta)=-2 A\arctan\left(A (1+\alpha)\tan \frac{\theta}{2} \right)
 \label{rcdef}
 \eeqa
 in the range $\theta \in [0,2\pi]$ and $u$ is a dummy integration variable. The letter $i$ denotes the imaginary unit.
 A calculation of the associated conformal factor $\lambda(\theta)$ is presented in Appendix \ref{app1} and given by the following expression
 \begin{equation}
    \lambda=\frac{R-r\cos{\theta}}{|\zeta|}. \label{cnfactor}
\end{equation}
  In terms of the complex co-ordinates, the hydrodynamic Green's function $G_H$ (or equivalently the stream function) has the following structure (we refer to Green and Marshall Ref.\cite{grms} for details)
\begin{equation}
    G_H(\zeta,\zeta_j)=\frac{1}{2\pi}\log\left|P\left(\frac{\zeta}{\zeta_j}\right)\right|+\varsigma~(\eta)+\left(\frac{\log|\zeta_j|}{4\pi^2A}-\frac{1}{4\pi}\right)\log|\zeta|- \int_{0}^{\theta_j}\frac{du}{4\pi^2\alpha}\frac{\alpha(u+\pi)-\sin u}{\alpha-\cos u} \;\label{GH}\\
\end{equation}
where
\begin{eqnarray}
 P(\zeta)&=&(1-\zeta)\prod_{n\geq1}(1-\rho^n\zeta)(1-\rho^n\zeta^{-1})\nn\\
  \varsigma(\eta)&=&\frac{A}{2\pi^2}\Re\left[\text{Li}_2(c^{-1}\eta)\right]-\frac{1}{2\pi^2\alpha}\log |\eta-c|-\frac{1}{8\pi^2 A}(\log|\zeta|)^2\nn\\
  \eta&=&|\zeta|^{\frac{i}{A}}.\nn
  %\varsigma(\eta)&=&-2iA\int\frac{\xi(\eta)}{\eta}d\eta-\frac{1}{4\pi^2\alpha}\log(-c),\nn\\
   \label{G1}
  \end{eqnarray}
  where $c~=-\alpha-A^{-1}$ as defined before Eq.~(\ref{vsol}).
  Let us also note that  the last term of Eq.(\ref{GH}) is to be thought of as a function of $\theta_j$ or equivalently a function of  $|\zeta_j|$ via the conformal map Eq.~(\ref{conformalmap}). Also, $\Re$ denotes the real part, $P(\zeta)$ is the Schottky-Klein prime function (Ref.~\cite{crowdymarshall,crowdyskpaper,crowdybook} for the concentric annulas, $\rho<|\zeta|<1$ and $\text{Li}_2$ represents the di-logarithmic function.  The hydrodynamic Green's function defined in Eq.(\ref{GH}) is singular in the limit $\zeta \rightarrow \zeta_j$. Hence we regulate it by subtracting a term which is essentially the logarithm of the infinitesimal geodesic distance on the torus written in terms of the conformal factor and given by $\log\left[\lambda(\zeta_j) |\zeta -\zeta_j|\right]$. This term cancels the logarithmic singularity appearing in the first term of Eq.(\ref{GH}) and we are left with the regulated stream function $\psi$ given by
  \beqa
   \psi (\zeta_m)=\sum_{j\neq m}^N \Gamma_j G_H(\zeta_m,\zeta_j)+\frac{1}{2}\Gamma_m R(\zeta_m)
  \label{strm}
  \eeqa
  where the Robin function $R(\zeta_m)$ is given by
  \beqa
R(\zeta_m)=&\frac{\log\prod_{n\geq 1}\left[1-\rho^n\right]^2}{2\pi}+\varsigma(\eta_m)+\left[\frac{\log|\zeta_m|}{4\pi^2A}-\frac{1}{4\pi}\right] \log|\zeta_m|- \int_{0}^{\theta_m}\frac{du}{4\pi^2\alpha}\frac{\alpha(u+\pi)-\sin u}{\alpha-\cos u}\nn\\&-\frac{1}{2\pi}\log\left[\lambda(\zeta_m)|\zeta_m|\right]
  \label{robin}
  \eeqa
  In the above expression, the last term in the first line of Eq.(\ref{robin}) is to be thought of as a function of $|\zeta_m|$ via the conformal map Eq.(\ref{conformalmap}). The vortex Hamiltonian constructed from the kinetic energy (see for example, Ref.\cite{boattod}) of the vortices is given in terms of the complex coordinate $\zeta$ as
  \begin{equation}
\mathcal{H}=-\frac{1}{2} \sum_{m=1}^N \sum_{j \neq m}^N \Gamma_m \Gamma_j G_{H}\left(\zeta_m, \zeta_j\right)-\frac{1}{2} \sum_{m=1}^N \Gamma_m^2 R\left(\zeta_m\right),
\label{hm}
\end{equation}
where the functions $G_H$ and $R$ are defined in Eq.~(\ref{GH}) and Eq.~(\ref{robin}) respectively.
  The Hamiltonian equations governing the dynamics of the m'th vortex  with coordinates ($\theta_m$,$\phi_m$) are described by the following equations:
\begin{eqnarray}
&r^2\left(\alpha-\cos \theta_m\right) \frac{\mathrm{d} \theta_m}{\mathrm{~d} t}=\mathrm{i} \sum_{j \neq m}^N \Gamma _j\left[\frac{K\left(\zeta_m / \zeta_j\right)-\overline{K\left(\zeta_m / \zeta_j\right)}}{4 \pi}\right]\nn \\
&r^2\left(\alpha-\cos \theta_m\right)^2 \frac{\mathrm{d} \phi_m}{\mathrm{~d} t}=  \sum_{j \neq m}^N \Gamma_j\left[\frac{K\left(\zeta_m / \zeta_j\right)+\overline{K\left(\zeta_m / \zeta_j\right)}}{4 \pi}+\frac{\alpha \theta_m-\sin \theta_m}{4 \pi^2 \alpha}+\frac{r_c\left(\theta_j\right)}{4 \pi^2 \mathcal{A}}-\frac{1}{4 \pi}\right]\nn \\
& +\Gamma_m\left[\frac{\alpha \theta_m-\sin \theta_m}{4 \pi^2 \alpha}+\frac{r_c\left(\theta_m\right)}{4 \pi^2 \mathcal{A}}+\frac{1}{4 \pi} \sin \theta_m\right].
\label{dyneq}
\end{eqnarray}
Here $\Gamma_j$ is a constant related to the circulation of the j'th vortex and 
 \begin{equation}
K(\zeta)= \frac{1}{1-\zeta}- \frac{1}{2  \pi A}~ \psi_\rho \left (\frac{\log \zeta}{2 \pi A} \right) +\frac{1}{2  \pi A} ~ \psi_\rho \left ( -\frac{\log \zeta}{2 \pi A} \right) 
\label{Kdef}
\end{equation}
The $q$-digamma function $\Psi_q(z)$ is the logarithmic derivative of the $q$-gamma function Ref.~\cite{qg0,qg1,qg2,qg3,qg4,qg5}
$$
\psi_q(z)=\frac{1}{\Gamma_q(z)} \frac{\partial \Gamma_q(z)}{\partial z}
=-\ln (1-q)+\ln q \sum_{n=0}^{\infty} \frac{q^{n+z}}{1-q^{n+z}}.
$$
 Here we evaluate it at $q=\rho$  and $z=\pm \frac{\log \zeta}{2 \pi A}$ to obtain Eq.~(\ref{Kdef}). 
Let us note that the dynamical equations Eq.~(\ref{dyneq}) have translational invariance along the $\phi$ direction but not along $\theta$  due to varying curvature. During the vortex evolution, the quantity 
 \begin{equation}
     C=\sum_{m=1}^{N}\Gamma_m(\alpha\theta_m - \sin\theta_m)
     \label{cnst}
 \end{equation}is invariant in time, please see Appendix Sec.\ref{app2} for details. The symplectic structure for $N$ vortices on the torus is given by 
$\omega_N=\sum_{m=1}^N \Gamma_m(\alpha-\cos\theta_m)\,d\theta_m\wedge d\phi_m$, 
and the conserved quantity takes the form 
$C=\sum_{m=1}^N \Gamma_m(\alpha\theta_m-\sin\theta_m)$. 
There exists a vector field $Y=-\sum_{m=1}^N \partial_{\phi_m}$ such that $dC=\iota_Y\omega_N$, 
which shows that $C$ is the momentum map associated with simultaneous $\phi$-rotations of all vortices. 
Consequently, the Hamiltonian vector field of $C$ is $X_{C}=-Y=\sum_{m=1}^N \partial_{\phi_m}$, whose flow simply advances all longitudes as $\phi_m(t)=\phi_m(0)+t$ while keeping $\theta_m$ fixed. 
Thus $C$ generates a uniform rigid rotation of the vortex configuration and Poisson-commutes with the Hamiltonian, confirming its interpretation as a conserved momentum map for global $\phi$-rotations on the torus.\\\\ We will also be interested in a  quantity ``$D$" for N-vortex clusters which is a measure of the sum of the inter-vortex distances
 \beqa
 D= \sum_{i \ne j}^N d_{ij}
 \label{ddef}
 \eeqa
 where $d_{ij}$ is the Euclidean distance between i-th and j-th vortex on the torus.
 For the purpose of numerical integration we choose $r=0.5$ and $R=1$ such that the parameter $\alpha=2$. The objective of the rest of the paper is to integrate Eq.~(\ref{dyneq}) with these set of parameters and investigate one and two vortex configurations in Sec.~(\ref{onetwovortex}), followed by a study of dynamics of vortex clusters on the torus in Sec.~(\ref{multivortex}). In all the following analysis, we utilize the well-tabulated q-digamma functions $\Psi_q(z)$.

\section{One and two vortex configurations}
\label{onetwovortex}
\begin{figure}[h!]
\begin{tabular}{lcccccccc}
\includegraphics[width=0.23\linewidth]{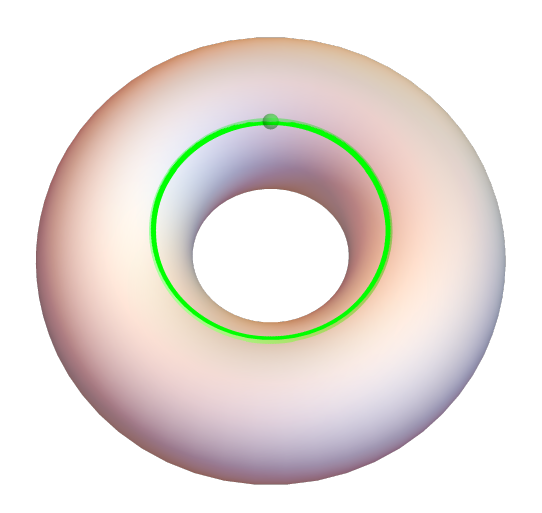}&&
\includegraphics[width=0.23\linewidth]{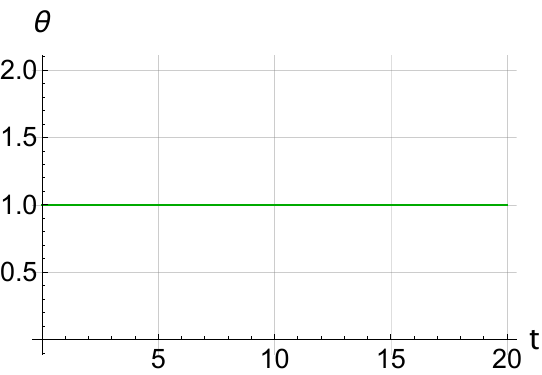}&&
\includegraphics[width=0.23\linewidth]{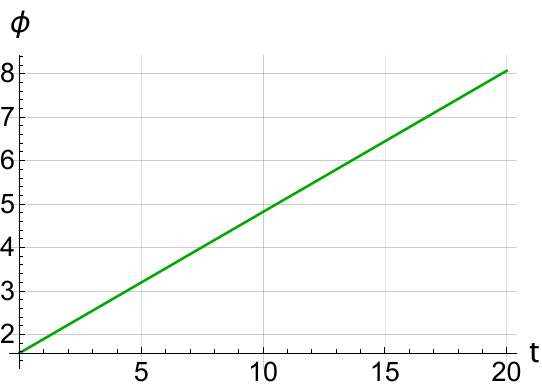}&&
\end{tabular}
 \caption{ Motion of a single vortex ($\Gamma=+1$)  with initial position $(\theta,\phi)=(1,\pi/2)$: we show the 3D plot of the trajectory on the extreme left, with a green dot marking the initial location of the vortex with the subsequent trajectory marked by a green curve. In the middle and right, we show the time evolution of vortex coordinates $\theta$ and $\phi$.}
 \label{singlevortex} 
\end{figure}
\begin{figure}[h!]
\begin{tabular}{ccc}
\includegraphics[width=0.23\linewidth]{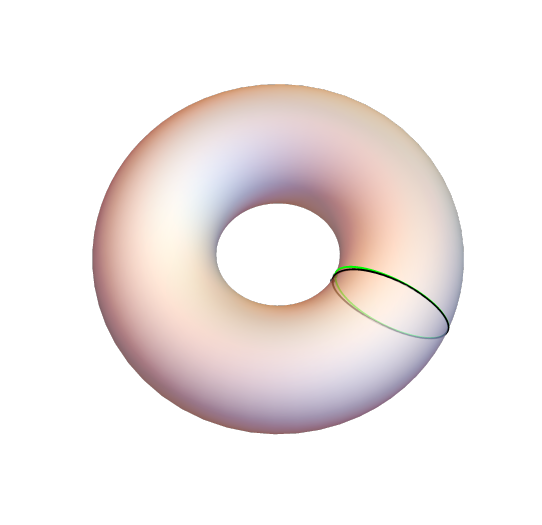}&~~~~
\includegraphics[width=0.23\linewidth]{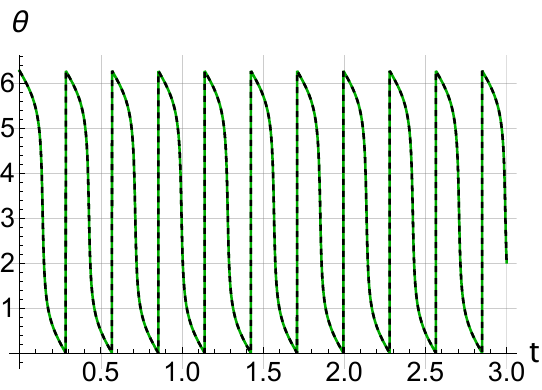}
\includegraphics[width=0.23\linewidth]{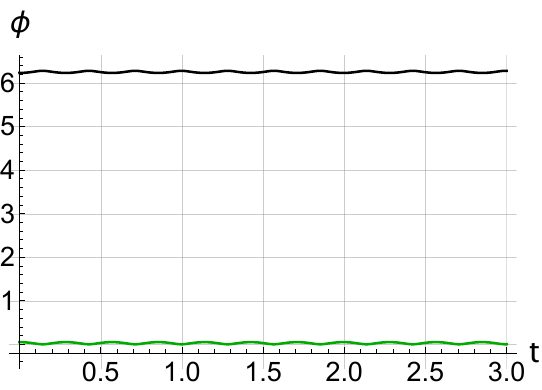}\\
\includegraphics[width=0.2\linewidth]{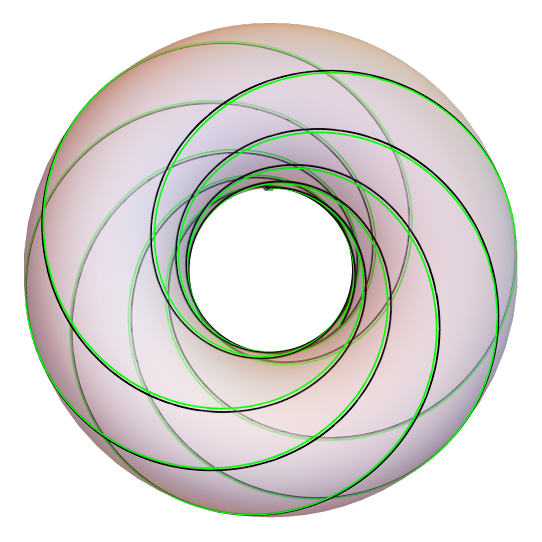}&
\includegraphics[width=0.23\linewidth]{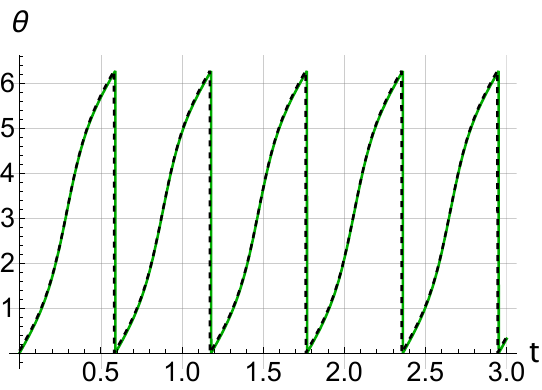}
\includegraphics[width=0.23\linewidth]{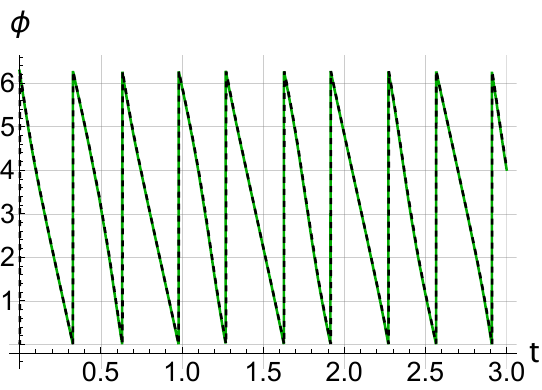}\\

\includegraphics[width=0.23\linewidth]{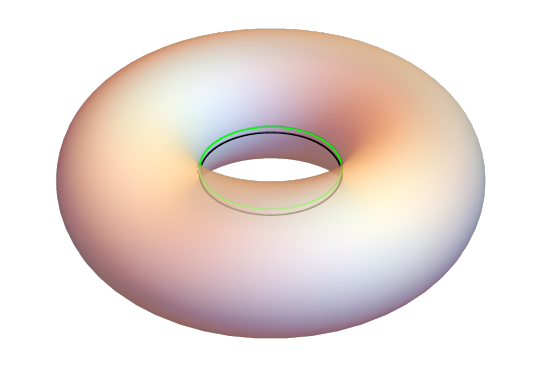}&
\includegraphics[width=0.23\linewidth]{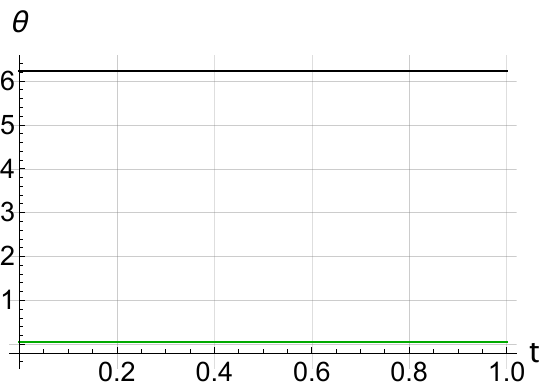}
\includegraphics[width=0.23\linewidth]{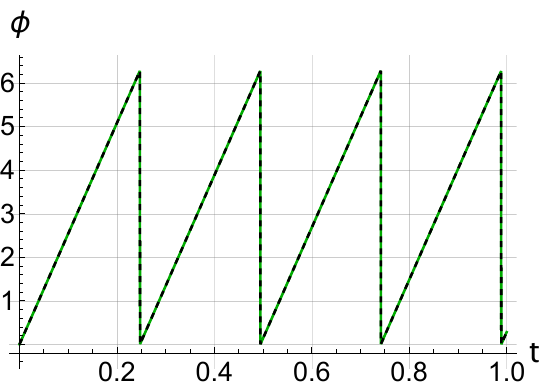}\\

\includegraphics[width=0.2\linewidth]{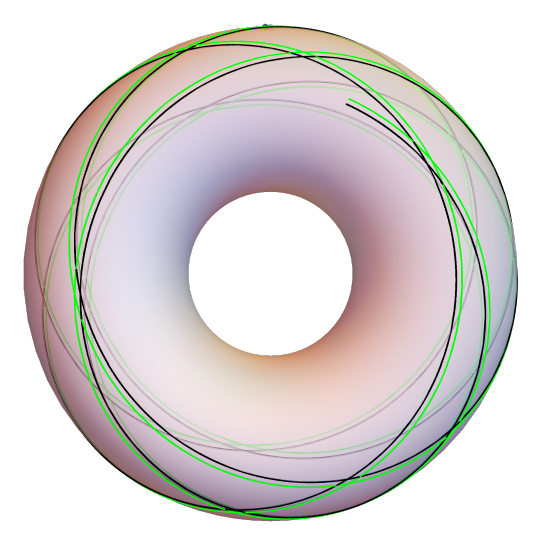}&
\includegraphics[width=0.23\linewidth]{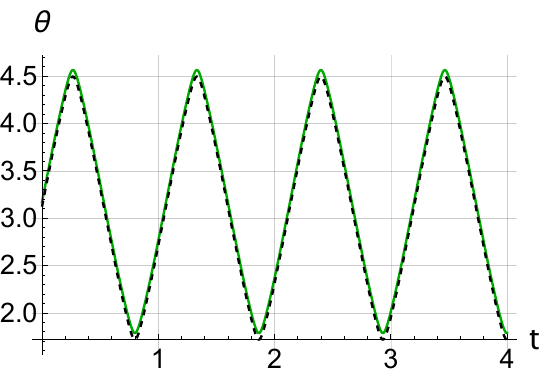}
\includegraphics[width=0.23\linewidth]{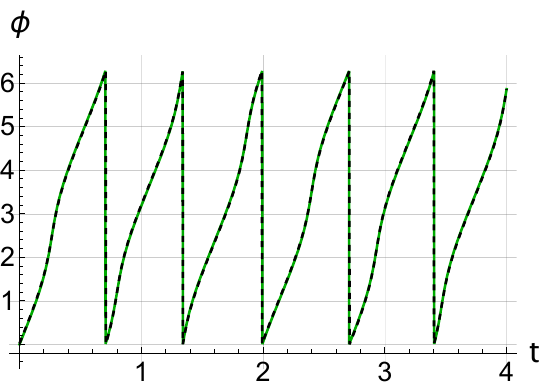}\\

\includegraphics[width=0.23\linewidth]{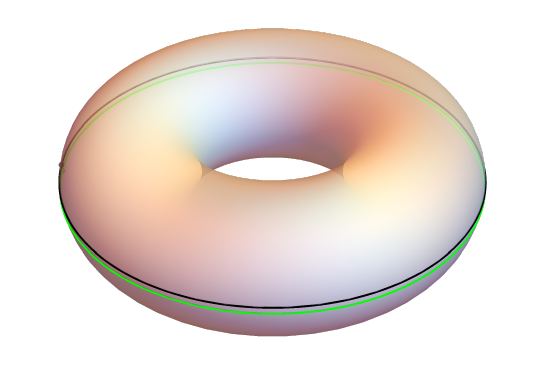}&
\includegraphics[width=0.23\linewidth]{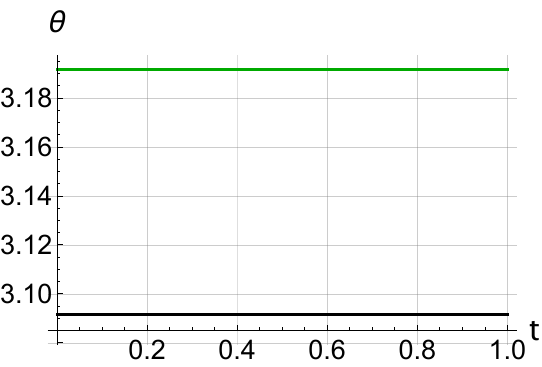}
\includegraphics[width=0.23\linewidth]{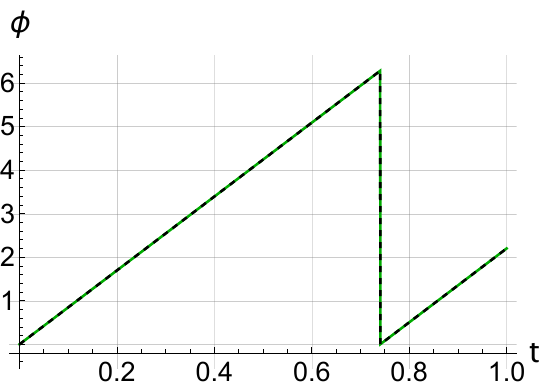}\\
    \end{tabular}\\
\caption{Five classes of geodesics on the torus: In all figures, the green dots represent location of the ``+" vortex at different instants of time, while ``-" vortex is indicated by black dots. 1st row : meridional geodesic generated by a dipole configuration separated along $\phi$, with $(\theta_1,\phi_1,\theta_2,\phi_2) =(0,0.05,0,-0.05)$, 2nd row: ``unbounded geodesics" alternately crossing the inner and outer equators, with initial conditions $(\theta_1,\phi_1,\theta_2,\phi_2)=(0,0,0.05,0.05)$,  3rd row: geodesic along the inner equator, with initial conditions $(\theta_1,\phi_1,\theta_2,\phi_2) =(0.05,0-0.05,0)$, 4th row: ``bounded" geodesics"  crossing outer equator multiple times but not the inner equator, with initial conditions $(\theta_1,\phi_1,\theta_2,\phi_2) =(3.17,0-3.12,0.02)$,  5th row: geodesic along the outer equator, with initial conditions $(\theta_1,\phi_1,\theta_2,\phi_2) =(\pi+0.05,0,\pi-0.05,0)$. In each row, the dipole trajectory on the torus is shown on the left, followed by $\theta$ and $\phi$ variation of both dipoles. The plots of $\theta$ and $\phi$ are restricted to periodic domains. }
     \label{vd}
\end{figure}
\begin{figure}[h!]
\begin{tabular}{lcccccccc}
\includegraphics[width=0.22\linewidth]{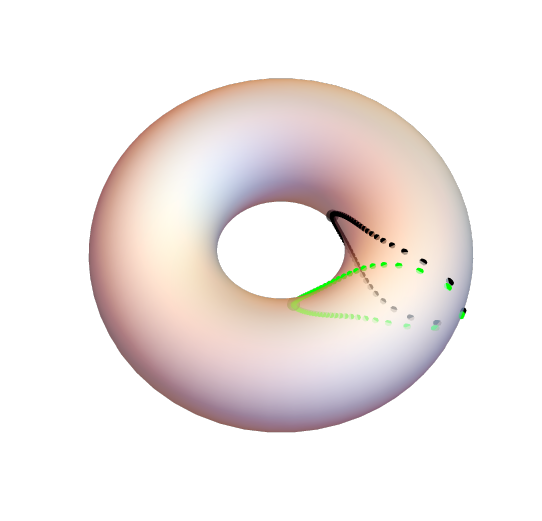}&&
\includegraphics[width=0.22\linewidth]{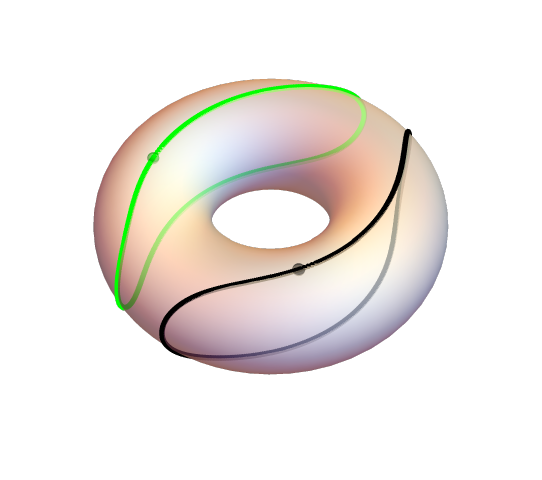}&&
\includegraphics[width=0.22\linewidth]{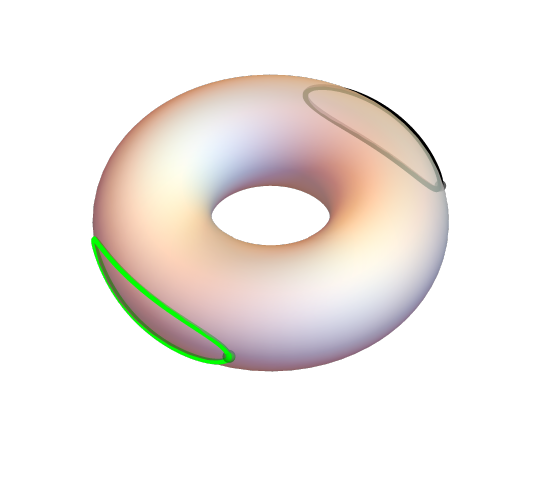}&&
\includegraphics[width=0.22\linewidth]{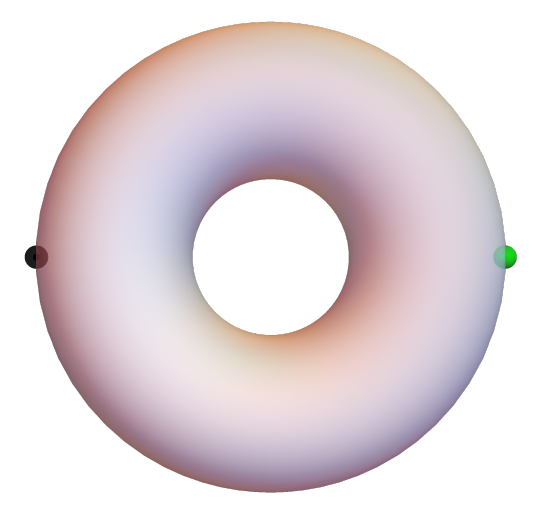}&&
\end{tabular}
 \caption{ Vortex dipole configuration symmetrically placed along the inner equator around the $\phi = 0$ meridian. In the extreme right configuration, the vortex and anti-vortex are diametrically opposite, resulting in a fixed equilibrium. The initial vortex positions are marked by a green dot for the vortex and a black dot for the anti-vortex, with trajectories shown in matching colors. From left to right, the initial conditions are $(\theta_1,\phi_1,\theta_2,\phi_2) = (0, 1, 0, -1)$, $(\pi/2, 1, \pi/2, -1)$, $(\pi, 1, \pi, -1)$, and $(\pi, \pi/2, \pi, -\pi/2)$, respectively. The rightmost configuration (diametrically opposite) differs from the behavior in quantum superfluids, where it exhibits nontrivial dynamics, as noted in Ref.~\cite{fetter}.}
 \label{vdd} 
\end{figure}
\begin{figure}[h!]
\begin{tabular}{lcccccccc}
\includegraphics[width=0.18\linewidth]{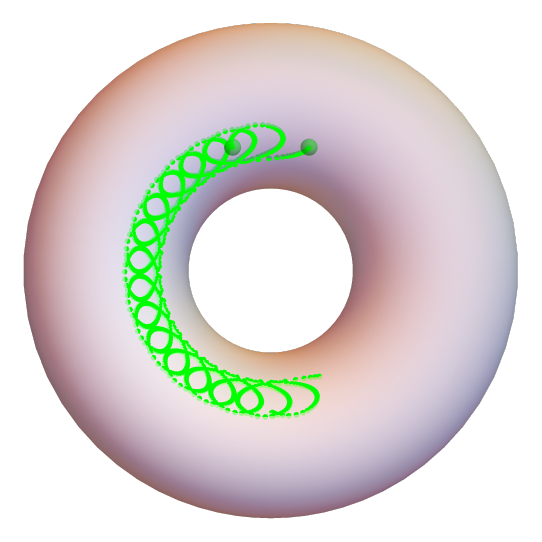}&&
\includegraphics[width=0.22\linewidth]{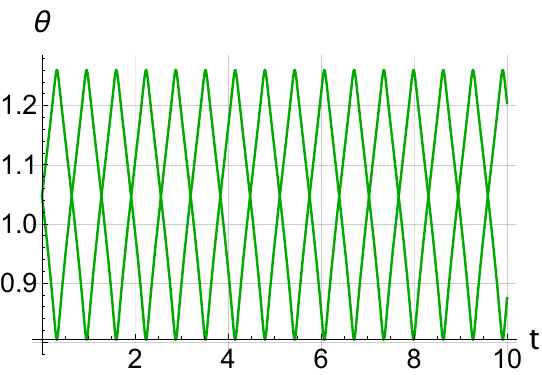}&&
\includegraphics[width=0.22\linewidth]{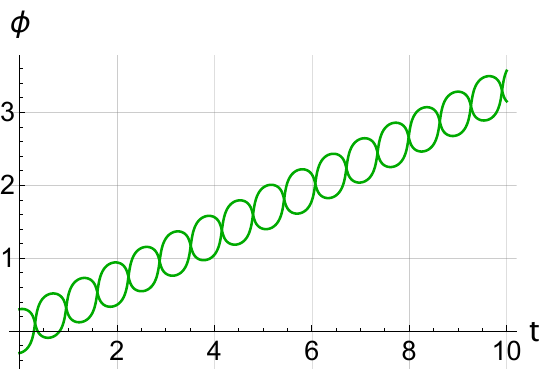}&&
\includegraphics[width=0.22\linewidth]{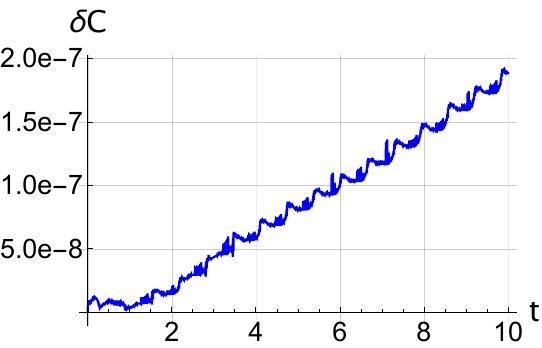}&&
\end{tabular}
 \caption{ Chiral vortex pair: From left to right, we show the 3D plot, $\theta$ vs time, $\phi$ versus time and numerical errors in C defined in Eq.(\ref{cnst}) of main text. The initial locations are $(\theta_1,\phi_1, \theta_2,\phi_2) =(\pi/3,0.3,\pi/3,-0.3)$.}
 \label{vc} 
\end{figure}
\begin{figure}[h!]
\begin{tabular}{ccccccccc}
\includegraphics[width=0.16\linewidth]{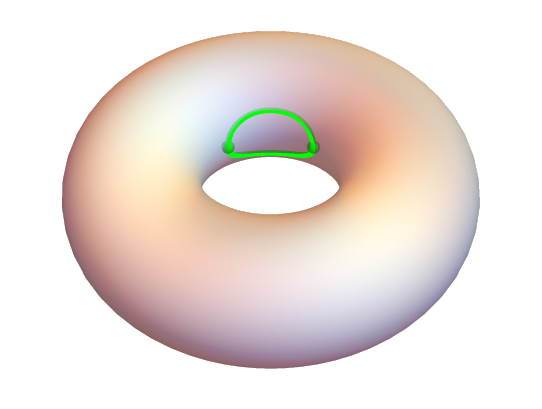}&&
\includegraphics[width=0.18\linewidth]{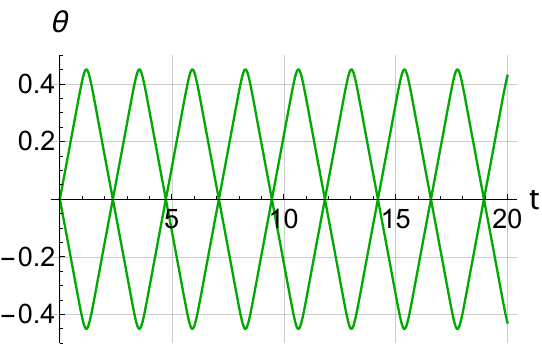}&&
\includegraphics[width=0.18\linewidth]{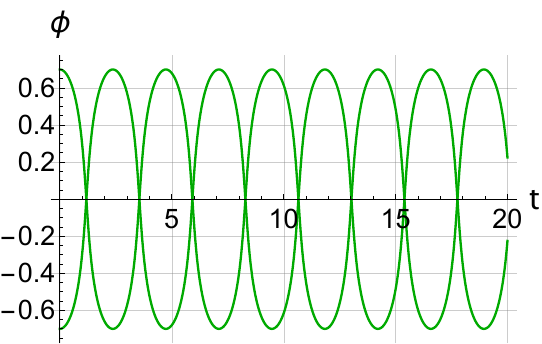}&&
\includegraphics[width=0.18\linewidth]{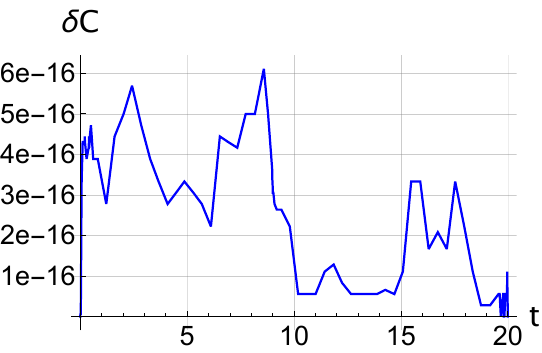}&&
\includegraphics[width=0.18\linewidth]{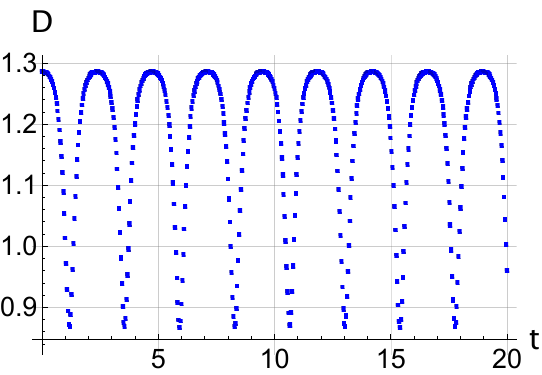}\\
\includegraphics[width=0.18\linewidth]{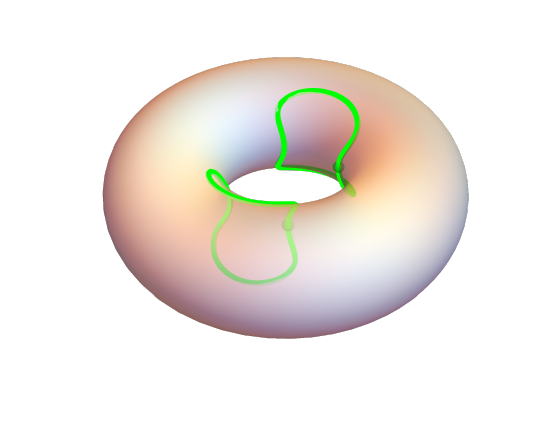}&&
\includegraphics[width=0.18\linewidth]{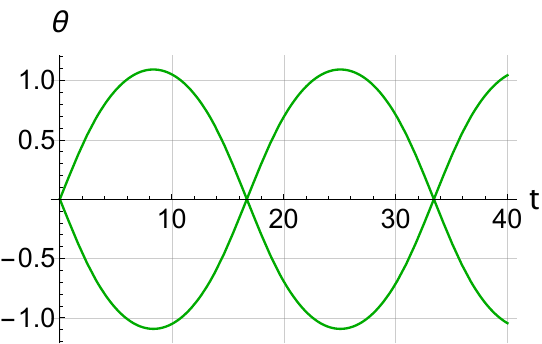}&&
\includegraphics[width=0.18\linewidth]{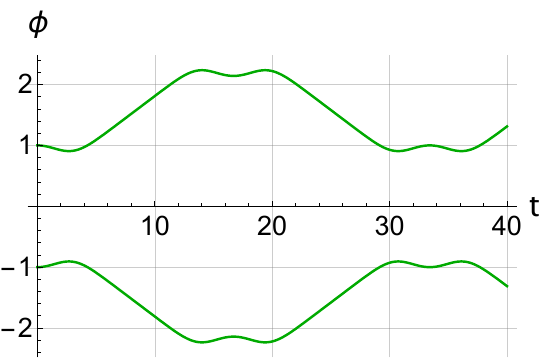}&&
\includegraphics[width=0.18\linewidth]{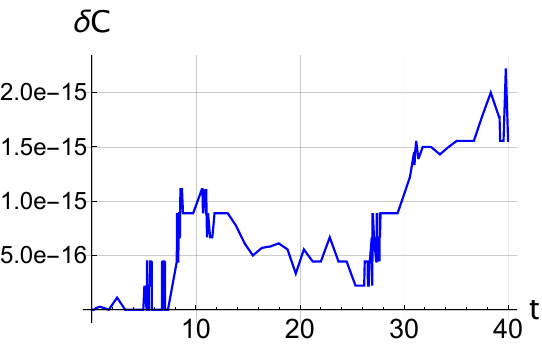}&&
\includegraphics[width=0.18\linewidth]{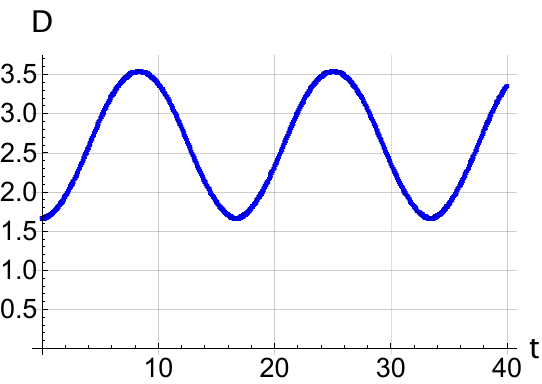}\\
\includegraphics[width=0.18\linewidth]{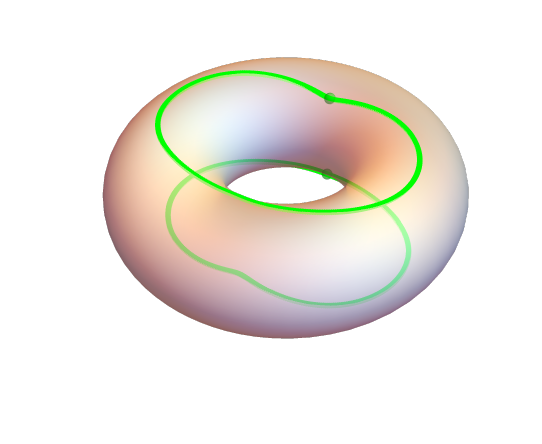}&&
\includegraphics[width=0.18\linewidth]{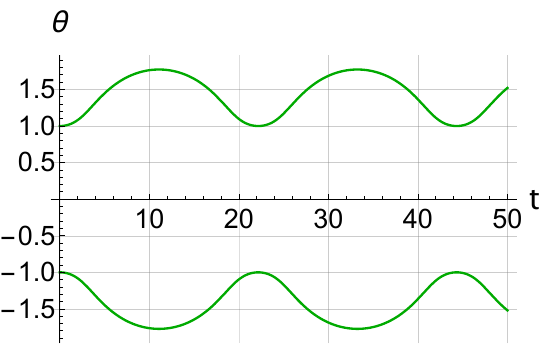}&&
\includegraphics[width=0.18\linewidth]{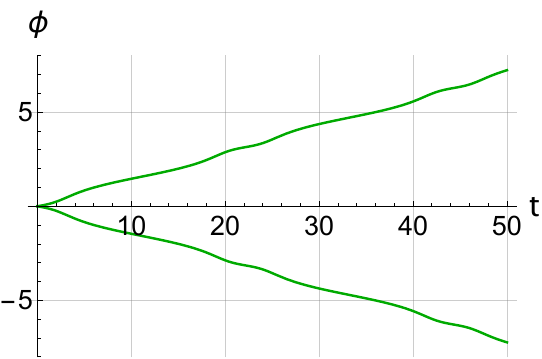}&&
\includegraphics[width=0.18\linewidth]{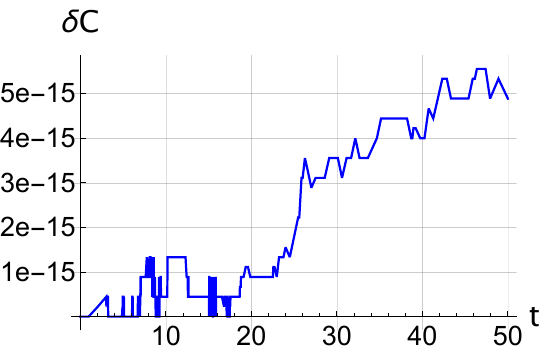}&&
\includegraphics[width=0.18\linewidth]{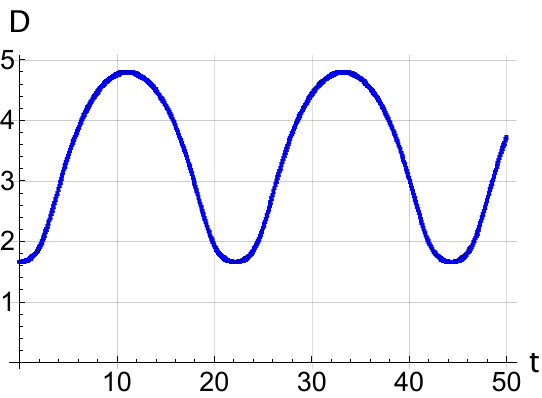}
\end{tabular}
 \caption{Top row: single-loop chiral vortex pair. Middle row: double-loop chiral vortex pair. Bottom row: another double-loop configuration. In each row, from left to right, we show the 3D trajectory, $\theta(t)$, $\phi(t)$, numerical error in $C$ defined in Eq.~(\ref{cnst}) of the main text, and the inter-vortex distance. The green dots in the 3D plots indicate the initial positions of the vortex pair, and the green curves represent their trajectories. The initial conditions are $(\theta_1,\phi_1,\theta_2,\phi_2) = (0, 0.7, 0, -0.7)$, $(0, 1, 0, -1)$, and $(1, 0, -1, 0)$, respectively.}
 \label{vcothers} 
\end{figure}

In this section, we explore one and two vortex configurations, both as a consistency check of our formulation involving  $q$-digamma functions with Ref.~\cite{grms,sakajo2016} as well as deriving some interesting insights on the fundamental vortex interactions on the torus.\\\\
\textbf{Single vortex}: The first interesting dynamics worth highlighting is that of a single vortex moving on the torus due to the \textit{self-force} of geometric origin, that arises from the Robin function Eq.~(\ref{robin}), leading to the \textit{self-drift} terms in the second equation of Eq.~(\ref{dyneq}). This results in a motion of the single vortex along the toroidal $\phi$ direction.  with constant speed, devoid of any $\theta$ variations as expected from the dynamical equations. This is illustrated in Fig.~(\ref{singlevortex}). Let us note that such a drift of the single vortex in absent in flat and spherical domains due to symmetry considerations, see for example Ref.~\cite{rs2021}. Moreover, from Eq.(\ref{dyneq}), we find that the drift term on the torus vanishes at the inner and outer equators corresponding to $\theta=0$ (negative curvature) and $\theta=\pi$ (positive curvature) respectively.\\\\
\textbf{Opposite sign (Vortex dipole) }: Next we focus on two-vortex dipole configuration, where we have two closely situated vortices having equal and opposite sign, also known as ``vortex-antivortex" configuration.  As shown in Fig.~(\ref{vd}), the dipole  traces out geodesic curves on the torus during it's motion, consistent with Kimura's conjecture, Ref.~(\cite{kimura,boattok,boattod,khesin2024}), now extended to surfaces of variable curvature like the torus. Depending on the initial conditions, we observe five distinct classes of geodesics. Apart from the inner and outer equators and the meridional geodesics, we have a class of geodesics which alternately cross the inner and outer equators (unbounded geodesics), while the fifth class consists of geodesics that never cross the inner equator and remain bounded in a band around the outer equator (bounded geodesics). One can also have a vortex dipole configuration where the vortex and antivortex are not necessarily closely spaced, but symmetrically placed along the inner equator around the $\phi=0$ meridian. This is illustrated in Fig.~\ref{vdd}. A key feature of these configurations is that the vortex and antivortex move symmetrically around $\phi=0$ meridian in such situations, such that $\phi_1+\phi_2=0$ at all times and $D_2$ defined in Eq.~(\ref{ddef}) exhibits periodic oscillations in time. The only exception is when the vortex and antivortex are  in diametrically opposite locations, where there is no motion and we have a fixed equilibrium. It is worth mentioning that this is distinct from the vortex dipole dynamics in toroidal superfluid films where additional quantum interaction (arising from single valued-ness of the condensate wavefunction)  gives rise to an extra term
Ref.~(\cite{fetter}), which leads to non-trivial dynamics in the diametrically opposite configuration, see Fig.(4c) of Ref.~(\cite{fetter}). However, we observe similar dynamics for all other dipole configurations in the classical fluid when compared with the quantized vortex dipoles in the toroidal superfluid case.\\\\
\textbf{Model comparisons:} Let us summarize the results in relation to existing literature. Firstly, the equations of motion Eq.~(\ref{dyneq}) show that in the ``thin" tori (flat cylinder) limit of large $\alpha$, the self-drift terms become subdominant, consistent with cylinder vortex dynamics Ref.~(\cite{tokieda, rs2025}). Secondly, we observe vortex dipoles moving along five classes of geodesics of the torus, consistent with Kimura's conjecture Ref.~(\cite{kimura}), extended to surfaces of generic curvature, Ref.~(\cite{ boattod,khesin2024}). Apart from checking consistency with the configurations discussed in Ref.~(\cite{sakajo2016, sakajo2019}), we also recover the dynamics described in Ref.~(\cite{fetter}) in the absence of the additional quantum interaction term. This creates distinct features for diametrically opposite vortex configurations, as reported in Ref.~(\cite{fetter}), consistent with our findings.
\\\\
\textbf{Same sign (chiral vortex pair)}:
The motion of two vortices with identical vortex strengths exhibits remarkably different dynamics compared to the vortex dipole. The generic motion in any curvature region of the torus is that of two vortices orbiting each other, while drifting along the toroidal $\phi$ direction, see Fig.~(\ref{vc}). This toroidal drift is absent in flat and spherical domains, see for example Ref.~(\cite{newton, rs2021}). However, the shape of the orbit as well as the number of loops depend both on the initial separation and on the curvature. This is illustrated in Fig.~(\ref{vcothers}) where we observe two vortices orbiting in a single loop or two separate loops depending on the initial condition. This transition from one-loop to two-loop can be easily achieved by starting with a closely spaced vortex pair in the inner equator and gradually increasing the separation along the toroidal or poloidal direction.

\section{Vortex Clusters}
\label{multivortex}
\begin{figure}[h!]
\begin{tabular}{ccc}
\includegraphics[width=0.25\linewidth]{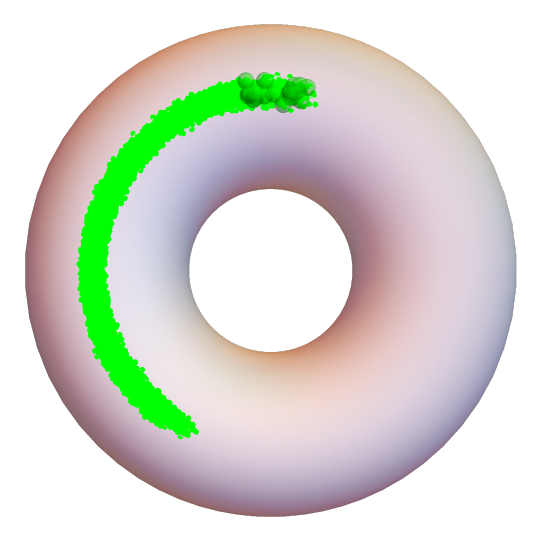}&
\includegraphics[width=0.3\linewidth]{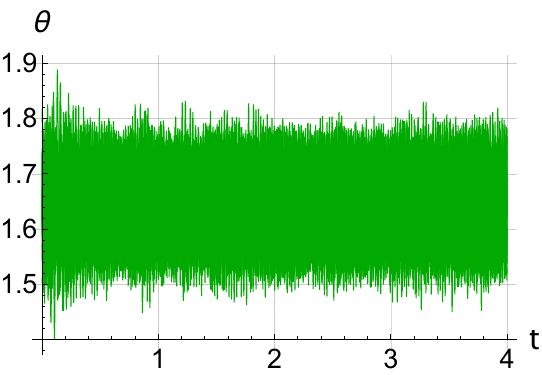}
\includegraphics[width=0.3\linewidth]{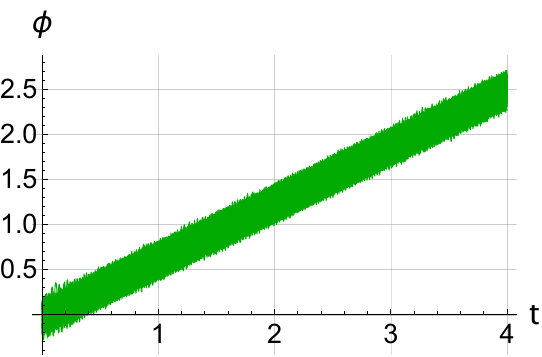}\\
\includegraphics[width=0.3\linewidth]{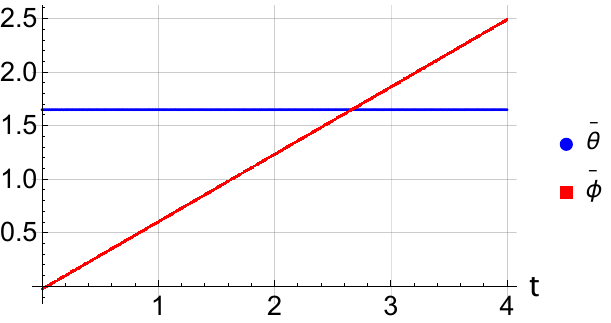}&
\includegraphics[width=0.3\linewidth]{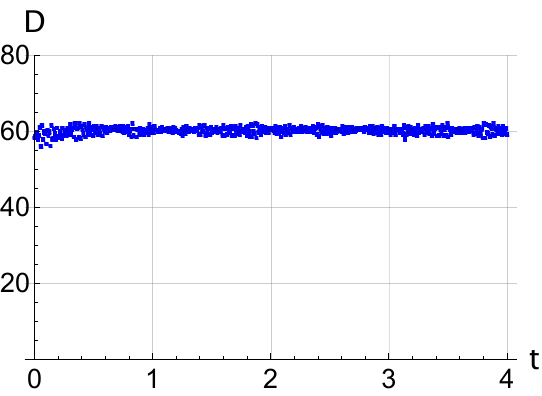}
\includegraphics[width=0.3\linewidth]{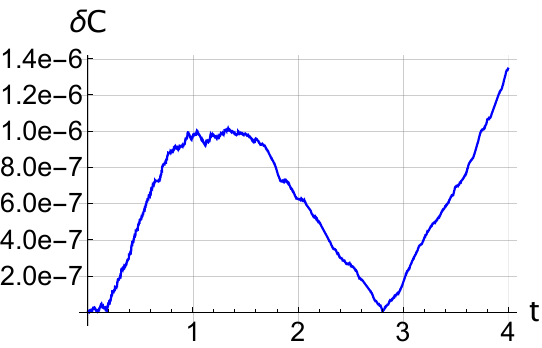}
 \end{tabular}\\
\caption{Time evolution of the chiral vortex cluster: Top row (left to right): 3D trajectory plot of the vortex cluster, $\theta$ vs. time, and $\phi$ vs. time for all vortices. Bottom row: Evolution of the mean coordinates $\bar{\theta}$ and $\bar{\phi}$, cluster size $D$ (sum of intervortex distances), and numerical error $\delta C$ [Eq.~(\ref{cnst})]. In the 3D plot, green dots mark the initial vortex positions and green curves their trajectories. Initial positions are uniformly distributed in $\theta \in (1.5, 1.8)$ and $\phi \in (-0.2, 0.2)$.}
     \label{mvc}
\end{figure}
\begin{figure}[h!]
\begin{tabular}{ccc}
\includegraphics[width=0.25\linewidth]{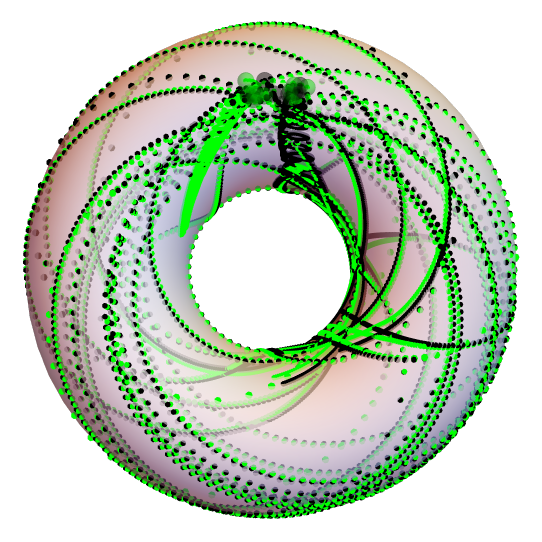}&
\includegraphics[width=0.3\linewidth]{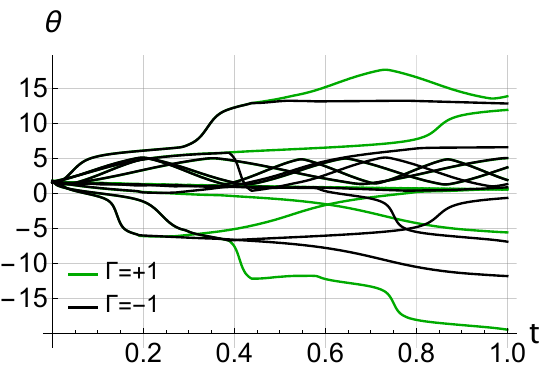}
\includegraphics[width=0.3\linewidth]{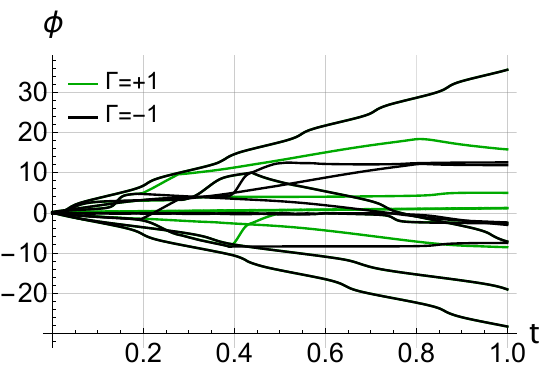}\\
\includegraphics[width=0.3\linewidth]{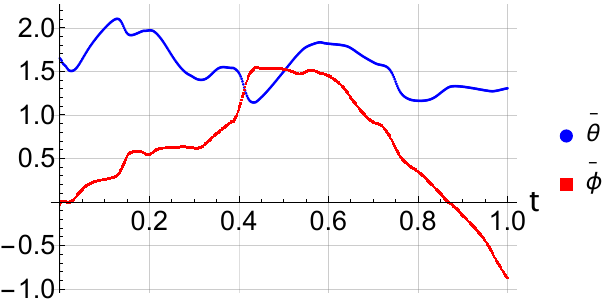}&
\includegraphics[width=0.3\linewidth]{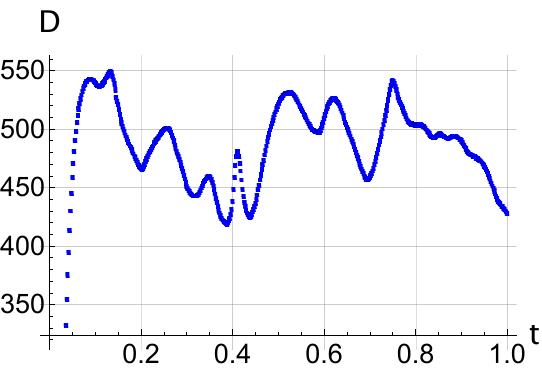}
\includegraphics[width=0.3\linewidth]{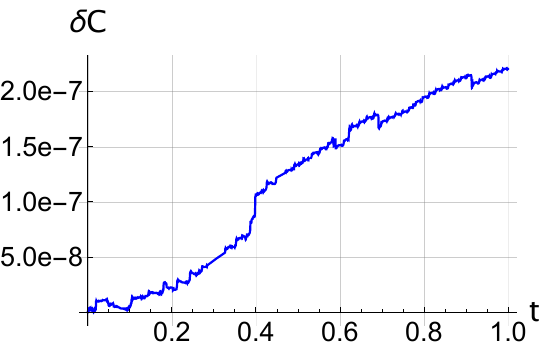}
 \end{tabular}\\
\caption{Time evolution of the achiral vortex cluster: Top row: 3D trajectory plot, $\theta$ vs. time, and $\phi$ vs. time for all vortices. Bottom row: $\bar{\theta}$ and $\bar{\phi}$ evolution, cluster size $D$, and $\delta C$ [Eq.~(\ref{cnst})]. In the 3D plot, green and black dots mark the initial positions of vortices and anti-vortices, respectively, with corresponding color-coded trajectories. Initial positions are uniformly distributed in $\theta \in (1.5, 1.8)$ and $\phi \in (-0.2, 0.2)$. }
     \label{mvac}
\end{figure}
\begin{figure}[h!]
\begin{tabular}{ccc}
\includegraphics[width=0.25\linewidth]{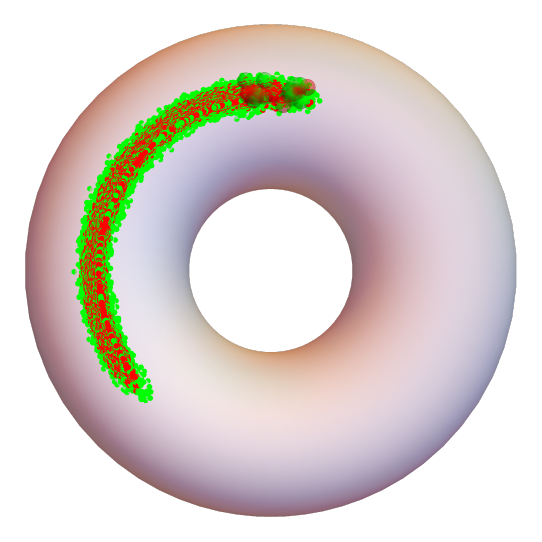}&
\includegraphics[width=0.3\linewidth]{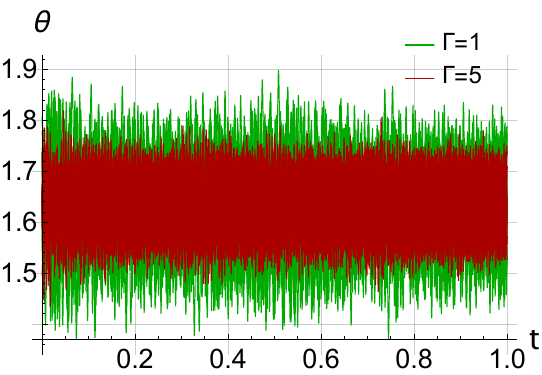}
\includegraphics[width=0.3\linewidth]{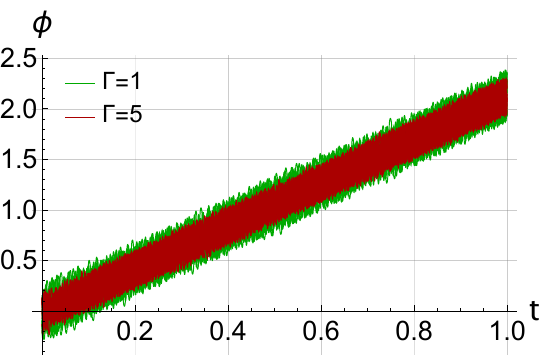}\\
\includegraphics[width=0.3\linewidth]{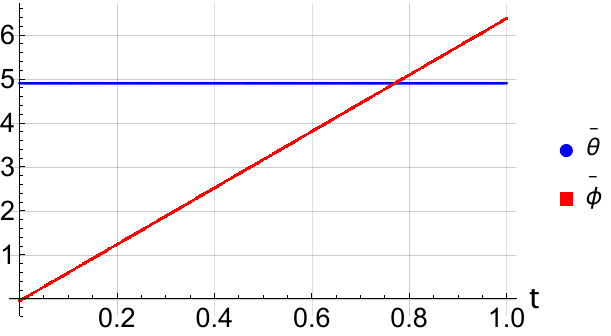}&
\includegraphics[width=0.3\linewidth]{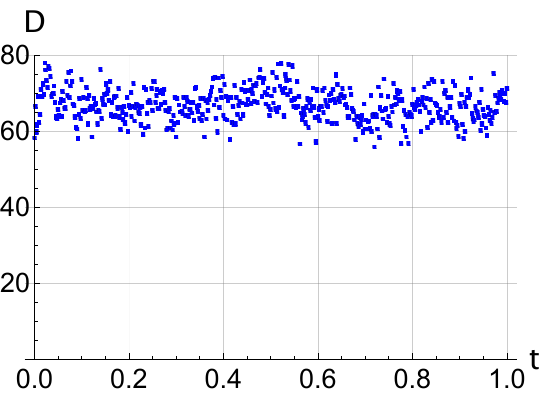}
\includegraphics[width=0.3\linewidth]{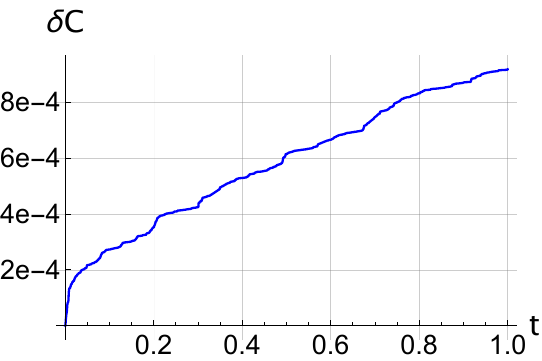}
 \end{tabular}\\
\caption{Time evolution of the fast–slow vortex cluster: Top row: 3D trajectory plot, $\theta$ vs. time, and $\phi$ vs. time for all vortices. Bottom row: $\bar{\theta}$ and $\bar{\phi}$ evolution, cluster size $D$, and $\delta C$ [Eq.~(\ref{cnst})]. In the 3D plot, green dots mark slow vortices ($\Gamma=+1$) and red dots mark fast vortices ($\Gamma=+5$), with matching trajectory colors. Initial positions are uniformly distributed in $\theta \in (1.5, 1.8)$ and $\phi \in (-0.2, 0.2)$.}
     \label{mvfs}
\end{figure}
\begin{figure}[h!]
\begin{tabular}{ccc}
\includegraphics[width=0.3\linewidth]{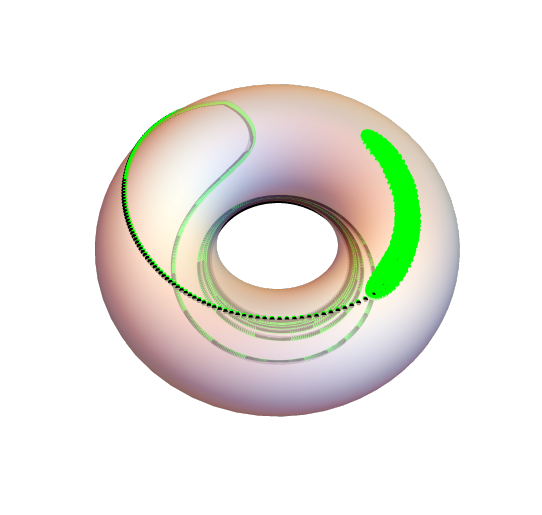}&
\includegraphics[width=0.3\linewidth]{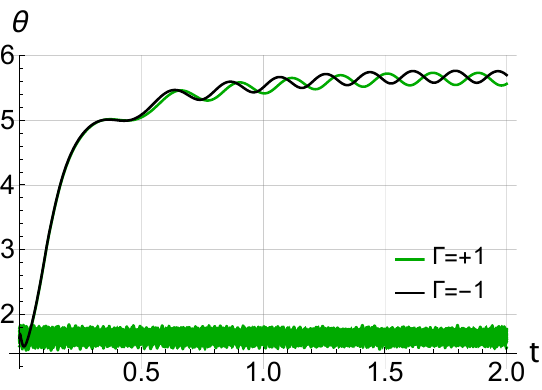}
\includegraphics[width=0.3\linewidth]{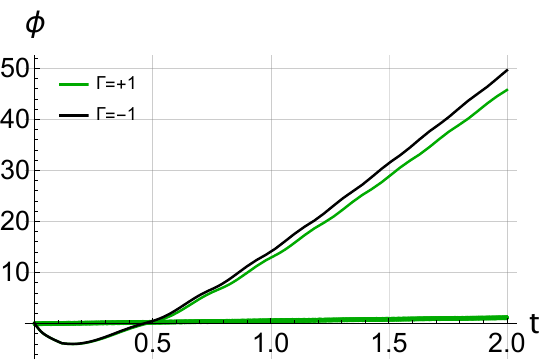}\\
\includegraphics[width=0.3\linewidth]{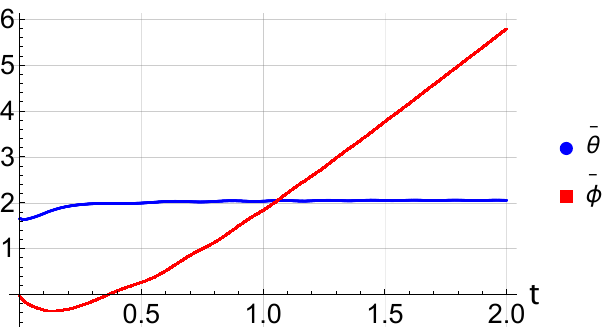}&
\includegraphics[width=0.3\linewidth]{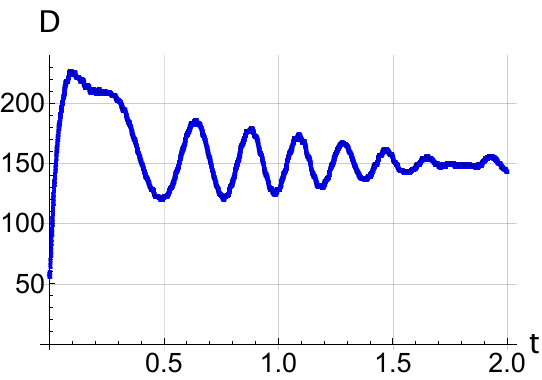}
\includegraphics[width=0.3\linewidth]{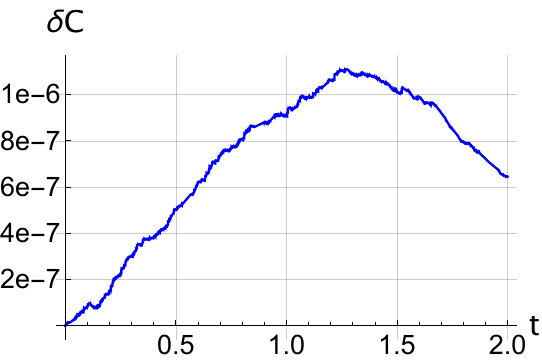}
 \end{tabular}\\
\caption{Time evolution of the chiral cluster with a single negative impurity: Top row: 3D trajectory plot, $\theta$ vs. time, and $\phi$ vs. time for all vortices. Bottom row: $\bar{\theta}$ and $\bar{\phi}$ evolution, cluster size $D$, and $\delta C$ [Eq.~(\ref{cnst})]. In the 3D plot, green dots mark ``+'' vortices ($\Gamma=+1$) and the black dot marks the single impurity ($\Gamma=-1$), with trajectories in matching colors. Initial positions are uniformly distributed in $\theta \in (1.5, 1.8)$ and $\phi \in (-0.2, 0.2)$.}
     \label{mvci}
\end{figure}
\begin{figure}[h!]
\begin{tabular}{ccc}
\includegraphics[width=0.25\linewidth]{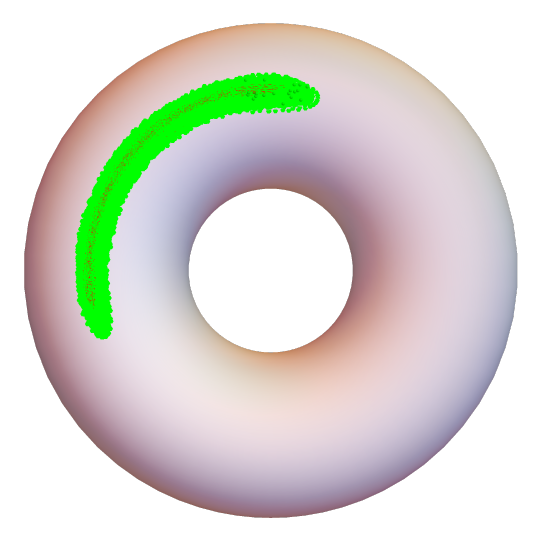}&
\includegraphics[width=0.3\linewidth]{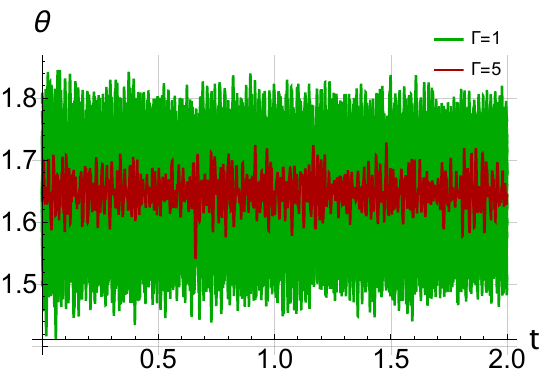}
\includegraphics[width=0.3\linewidth]{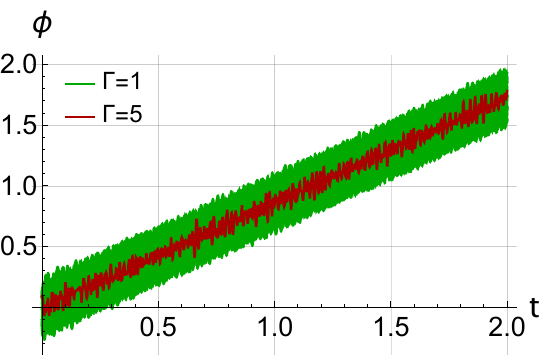}\\
\includegraphics[width=0.3\linewidth]{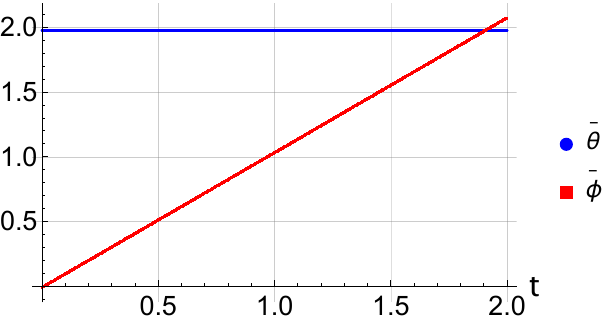}&
\includegraphics[width=0.3\linewidth]{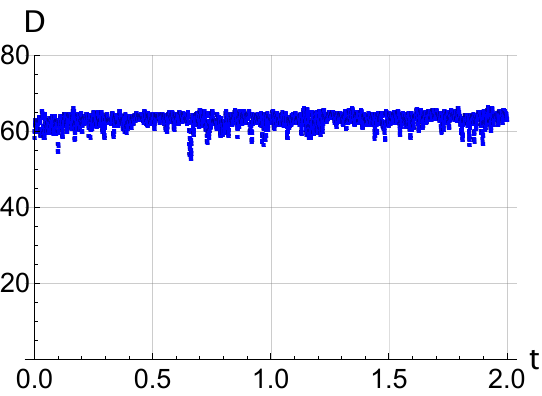}
\includegraphics[width=0.3\linewidth]{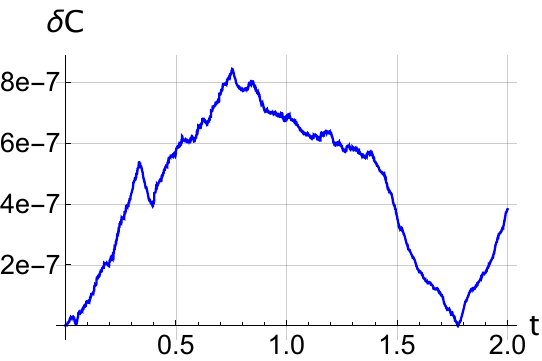}
 \end{tabular}\\
\caption{Time evolution of the chiral cluster with a single fast impurity: Top row: 3D trajectory plot, $\theta$ vs. time, and $\phi$ vs. time for all vortices. Bottom row: $\bar{\theta}$ and $\bar{\phi}$ evolution, cluster size $D$, and $\delta C$ [Eq.~(\ref{cnst})]. In the 3D plot, green dots mark slow vortices ($\Gamma=+1$) and the red dot marks the fast impurity ($\Gamma=+5$), with trajectories in matching colors. Initial positions are uniformly distributed in $\theta \in (1.5, 1.8)$ and $\phi \in (-0.2, 0.2)$.}
     \label{mvfi}
\end{figure}

We now examine the dynamics of a single vortex cluster on the torus. As a starting point, we consider a small cluster of 20 vortices and numerically integrate Eq.~(\ref{dyneq}) to track its time evolution. The vortex trajectories are computed using a fourth-order Runge–Kutta (RK4) scheme with adaptive time stepping. At each iteration, the time step is adjusted to ensure that the relative decrease in separation between any two approaching vortices does not exceed a prescribed threshold (0.05 or 0.1 in our simulations). In addition, the time step is bounded above (between 0.01 and 0.001, depending on the run) to keep the maximum local integration error per step in the range $10^{-7}$–$10^{-6}$. We investigate the five distinct configurations:
\begin{enumerate}
\item[(a)] Chiral cluster – 20 vortices of unit strength and identical circulation.
\item[(b)] Achiral/Neutral cluster – 10 vortices of positive unit circulation and 10 of negative unit circulation.
\item[(c)] Fast–slow chiral cluster – half with fast circulation ($\Gamma=5$), half with slow circulation ($\Gamma=1$).
\item[(d)] Chiral cluster with a single negative impurity – same as (a), but with one vortex replaced by a negative unit circulation vortex.
\item[(e)] Chiral cluster with a single fast impurity – same as (a), but with one vortex replaced by a fast circulation vortex ($\Gamma=5$).
\end{enumerate}
Fig.~(\ref{mvc}–\ref{mvfi}) present the corresponding simulations. In each figure, the top row (extreme left) shows a 3D visualization of the dynamics: initial vortex locations are marked with dots, and their trajectories are overlaid on the torus surface. Vortices with $\Gamma=1$ are shown in green, $\Gamma=-1$ in black, and $\Gamma=5$ in red. This is followed by the time evolution of $\theta$ and $\phi$ for all vortices. In the second row, the left panel shows the time evolution of the cluster-averaged position $(\bar{\theta},\bar{\phi})$, followed by the total inter-vortex distance $D$ [Eq.~(\ref{ddef})], and a plot of the numerical error. \\ \\
\textbf{(a) Chiral cluster}: The cluster is initialized as 20 closely spaced, randomly positioned vortices of identical circulation. Fig.~\ref{mvc} shows that the dynamics combines short-range inter-vortex interactions with a collective toroidal drift of the entire cluster, an effect absent in flat or spherical geometries. The total inter-vortex distance $D$ remains nearly constant at its initial value, indicating area-preserving interactions. This preservation confines the cluster to its initial ``curvature belt'' throughout the drift. Similar behavior occurs when the cluster is placed in other curvature regions of the torus.\\\\
\textbf{(b) Achiral/Neutral cluster}: The achiral cluster is constructed from a mixed population of 20 vortices, one half is made up of vortices of positive circulation and the other half of negative circulation, such that the total circulation of the cluster as a whole vanishes. Vortices are initialized at random locations within the cluster. This is displayed in Fig.~(\ref{mvac}) where we observe completely different dynamics. The achiral cluster quickly disintegrates and scatters all over the torus, showing unconfined dynamics.\\\\
\textbf{(c) Fast-slow cluster}: In this situation, see Fig.~(\ref{mvfs}), we have half the population built out of fast vortices and the other half slow. The vortices are again initialized at random locations within the cluster. During the time evolution, we again observe inter-vortex interactions and the collective toroidal drift of the chiral cluster. However, fast vortices move predominantly through the central core of the cluster, expelling the slow vortices to the outer periphery of the cluster. Let us note that the cluster expands a little in this situation, due to the expulsion of the slow vortices to the outer periphery. However, the cluster still maintains confined dynamics similar to a chiral cluster. Fast-slow clusters in other curvature regions of the torus show similar dynamics.\\\\
\textbf{(d) Chiral cluster with a single negative impurity}: In this situation, we start with the same chiral cluster configuration as (a) but replace one of the positive circulation vortices with a vortex of negative unit circulation (impurity). The cluster exhibits the collective toroidal drift along with inter-vortex interactions. However, during the course of its motion, the cluster often ejects a vortex dipole (vortex along with an anti-vortex). In Fig.~(\ref{mvci}) we show one such scenario where a dipole is ejected from the revolving cluster. In general, the ejected vortex dipole follows a geodesic on the compact torus for a while before again merging with the revolving bulk vortex cluster. A new vortex dipole again gets ejected from the cluster and the process continues.  We have also checked that the dynamics is similar for such a cluster in the other curvature belts of the torus.\\\\
\textbf{(e) Chiral cluster with a single fast impurity}: The cluster configuration is similar to the one described in (d), however the impurity is now of a fast circulation type ($\Gamma=5$) in the cluster of slow vortices ($\Gamma=1$). The cluster evolves similar to (c), with the impurity ie. the fast vortex moving through the central region of the cluster, see Fig.~(\ref{mvfi}) in contrast to the ejection of impurity from the cluster as seen in (d). We also checked that this evolution of the cluster remains same in all curvature belts of the torus.\\\\
Let us end this section with some comments on the parameter $\alpha \equiv R/r$ which we have set to 2 throughout this work. The parameter essentially provides a transition from the thin tori (cylinder limit) to the more interesting ''thick" tori regime ($\alpha \sim 1$).  From the equations of motion it is clear that as one increases $\alpha$, the self-drift terms of the vortices become subdominant compared to the other terms. Thus, the toroidal drift of the cluster decreases for thinner tori. However, a systematic study of the dynamics of vortex clusters with respect to variations in the $\alpha$ parameter merits a separate extensive study, which we plan to report in future.
\section{ Discussions}
\label{cncl}
In summary, we have explored the incompressible and inviscid fluid dynamics of vortices in a classical fluid film of toroidal shape, with emphasis on the time evolution of a single vortex cluster. The Hamiltonian formulation in terms of q-digamma functions reveals that the emergent dynamics of vortex clusters on the torus, although following the somewhat complicated dynamical equations Eq.~(\ref{dyneq}), can be constructed in terms of the fundamental motion observed with one and two vortex systems on the torus. In contrast to flat and spherical domains, a single vortex moves along the toroidal direction. A closely spaced vortex dipole moves along any one of five classes of geodesics, depending on initial conditions. Dipoles with larger separation exhibit constrained motion, differing from the quantum superfluid case where an additional quantized interaction term ensures single-valued condensate wavefunctions Ref.~\cite{fetter}, leading to distinct dynamics for diametrically opposite vortices. Regarding the dynamics of vortex clusters, we observe the general pattern that chiral clusters tend to interact in a manner which is area-preserving, while moving collectively along the toroidal direction. However, achiral clusters show unconfined dynamics and scatter throughout the torus. The collective toroidal drift of the chiral cluster is absent in spherical and flat domains. Impurities in chiral clusters evolve in a manner consistent with one and two vortex interactions. The dynamics of a fast-slow cluster is similar to chiral clusters, with fast vortices moving predominantly along the core region of the cluster, expelling the slow ones to the outer periphery. A chiral cluster with an impurity in the form of a single vortex of opposite sign also show similar behavior as a pure chiral cluster, with occasional ``jets" of dipoles leaving and re-entering the revolving cluster. The torus model thus provides a surprisingly simple yet elegant illustration of the topological implications of the underlying background geometry on vortex dynamics.\\\\
The work leverages the Hamiltonian structure on the torus in terms of q-digamma functions to provide a robust computational framework where the time evolution of the vortex clusters can be tracked easily. This extends the foundational work of Green and Marshall Ref.~(\cite{grms}) and Sakajo and Shimizu Ref.~(\cite{sakajo2016}) to the case of chiral clusters, achiral clusters, fast-slow clusters and clusters with a single impurity. Thus, it provides a platform to explore systems analogous to Ref.~(\cite{vsc}) in fluid domains of distinct curvature and topology. It also confirms and extends the Kimura conjecture for vortex dipoles Ref.~(\cite{kimura}) to surfaces of varying positive and negative curvature like the torus. Some of the robust predictions of the model, such as toroidal drift and area-preserving nature of chiral vortex clusters, separation of fast and slow dynamics during cluster revolution around the torus, as well as interesting behaviors of clusters with single impurities advance the existing literature on the subject. Although maintaining a homogeneous toroidal film in cold-atom experiments is experimentally challenging due to gravity, such traps have been achieved and proposed in microgravity platforms Ref.~\cite{trap1,trap2,iss1,iss2}, Our results should provide strong motivation for future experiments with vortex clusters in such systems.\\\\
 The work needs to be extended further in several directions to fully capture the rich dynamics we have seen with single vortex clusters. The natural line of investigation is to explore how two or more chiral/achiral/ fast-slow clusters in different curvature regions of the torus interact with each other.  It will also be interesting to perform similar investigation for achiral vortex clusters with quantized strengths in  toroidal superfluid films along the lines of Ref.~(\cite{fetter}). In particular, it will be worth exploring the signatures of the additional quantum interaction term on the dynamics of achiral vortex clusters reported in our work. It will also be interesting to explore the dynamics of vortex clusters in the viscous tubular fluid \textit{membrane} geometry of Ref.~(\cite{rs2025}), particularly the effects of membrane viscous stress and couplings to external solvents on the vortex dynamics constructed therein. Study of vortex clusters in more carefully constructed models incorporating harmonic fields (along the lines of Ref.~(\cite{h1,h2,h3}) will be reported in future.
\section{Acknowledgments}
We are very thankful to Haim Diamant, Naomi Oppenheimer, Takashi Sakajo, Arpan Saha, Ishaan Chaturvedi, Ishan Mata and Michael D. Graham. A.K.R is supported by an Institute fellowship from Birla Institute of Technology and Science, Pilani (Hyderabad Campus). R.S is supported by DST INSPIRE Faculty fellowship, India (Grant No.IFA19-PH231), NFSG and OPERA Research Grant from Birla Institute of Technology and Science, Pilani (Hyderabad Campus).
\section{Data Availability} The data that supports the findings of this study are available within the article.
 \begin{footnotesize}

\end{footnotesize}
\appendix
\section {Conformal factor $\lambda$ computation}
\label{app1}
In this section, we provide the details on the computation of the conformal factor $\lambda$, given by Eq.(\ref{cnfactor}) of the main text. The complex coordinates are
$\zeta=\exp{\left[{r_{c}(\theta)}+i\phi\right]}$ and $\bar{\zeta}=\exp{\left[{r_{c}(\theta)}-i\phi\right]}$ as given by Eq.(\ref{conformalmap}) of the main text. The function $r_c(\theta)$ is also defined in Eq.(\ref{rcdef}) of the main text. In terms of the complex coordinates, the metric of the toroidal surface is given by
\beqa
ds^2&=&\lambda^2 d\zeta d\bar{\zeta}
\label{metric2}
\eeqa
where
$d\zeta=\zeta\left(\frac{1}{\alpha-\cos{\theta}}d\theta+id\phi\right)$ and $
    d\bar{\zeta}=\zeta\left(\frac{1}{\alpha-\cos{\theta}}d\theta-id\phi\right)$. Now we note that
    \beqa
    d\zeta d\bar{\zeta}&=&\frac{\zeta \bar{\zeta}}{\left[R-r\cos{\theta}\right]^2}\left[r^2 d\theta^2+\left[R-r\cos{\theta}\right]^2 d\phi^2\right]\nn\\
    &=&\frac{|\zeta|^2}{\left[R-r\cos{\theta}\right]^2}\left[r^2 d\theta^2+\left[R-r\cos{\theta}\right]^2 d\phi^2\right].\nn
    \label{appeq2}
\eeqa
 This in turn implies that  
\beqa
   ds^2= \left[r^2 d\theta^2+\left[R-r\cos{\theta}\right]^2 d\phi^2\right]&=& \frac{\left[R-r\cos{\theta}\right]^2}{|\zeta|^2}d\zeta d\bar{\zeta}.
   \eeqa
   Comparing Eq.~(\ref{metric2}) and Eq.~(\ref{appeq2}), we find that $\lambda=\frac{\left(R-r\cos{\theta}\right)}{|\zeta|}$.
    
\section{Invariance of C }
\label{app2}
In this section, we show that the quantity ``C" represented by ~Eq.(\ref{cnst}) remains invariant in time. The time derivative of C can be evaluated as follows.
\begin{equation}
    \begin{split}
        \frac{dC}{dt}=\frac{d}{dt} \left(\sum_{m=1}^{N}\Gamma_m\left[\alpha\theta_m-\sin\theta_m\right] \right)\nn\\
        =\sum_{m=1}^{N}\Gamma_m\left[\alpha-\cos\theta_m\right]\;\frac{d\theta_m}{dt}.
        \end{split}
\end{equation}
Recalling ~Eq.(\ref{dyneq}) of the main text, the above derivative becomes
\begin{eqnarray}
       \left[\alpha-\cos\theta_m\right] \frac{d\theta_m}{dt} &=&\frac{i}{r^2}\sum_{j\neq m}^{N}\Gamma_j\frac{K(\zeta_m/\zeta_j)-\overline{K(\zeta_m/\zeta_j)}}{4\pi}.\nn
        \end{eqnarray}
    Since $\Gamma_m \Gamma_j$ is symmetric under interchange of indices, we have
        \begin{equation}
          \frac{dC}{dt}=\frac{i}{2}\sum_{m=1}^{N}\sum_{j\neq m}^{N}\frac{\Gamma_m \Gamma_j}{4\pi r^2}\left[K\left(\frac{\zeta_m}{\zeta_j}\right)+K\left(\frac{\zeta_j}{\zeta_m}\right)-\left[\overline{K\left(\frac{\zeta_m}{\zeta_j}\right)}+\overline{K\left(\frac{\zeta_j}{\zeta_m}\right)}\right]\right].\label{dc/dt}
  \end{equation}
  Plugging in the expression of $K$ from Eq.(\ref{Kdef}) of the main text ie. 
\beqa
K(\zeta)= \frac{1}{1-\zeta}- \frac{1}{2  \pi A}~ \psi_\rho \left (\frac{\log \zeta}{2 \pi A} \right) +\frac{1}{2  \pi A} ~ \psi_\rho \left ( -\frac{\log \zeta}{2 \pi A} \right) \nn
\eeqa
it is easy to see that the quantity within the brackets in the RHS of Eq.(\ref{dc/dt}) vanishes. Hence, C remains invariant in time, which is also ensured in our numerical analysis to a high degree of accuracy.
\section{Detailed derivation of the vortex dynamical equations}
\label{app3}
In this  section, we provide a road map leading to the derivation of the vortex dynamical equations described by Eq.~(\ref{dyneq}) of the main text. For more details, we refer the readers to the foundational works by Crowdy, Ref.~(\cite{crowdymarshall,crowdyskpaper}), Green and Marshall, Ref.~(\cite{grms}) and Sakajo and Shimizu, Ref.~(\cite{sakajo2016}). The key idea is to use the conformal map Eq.~(\ref{conformalmap}) and then use the known Green's function for the concentric annulus, Ref~(\cite{grms}). In terms of the complex coordinate $\zeta$, the vortex dynamical equations for a classical incompressible and inviscid fluid take the form (Ref.~(\cite{hally}))
\begin{equation}
    \frac{d\zeta_m}{dt}=2i\lambda^{-2}(\zeta_m,\bar{\zeta}_m)\frac{\partial\psi_m}{\partial\bar{\zeta}_m}\label{S2}
\end{equation}
where $\lambda$ is the conformal factor  and $\psi$ is the vortex streamfunction defined in Eq.~(\ref{cnfactor})  and Eq.~(\ref{strm}) of the main text respectively.  The final form of the vortex dynamical equations Eq.~(\ref{dyneq}) of the main text is in terms of coordinates $\theta$ and $\phi$  on the torus. Thus, we can use Eq.~(\ref{conformalmap}) of the main text to write the L.H.S of Eq.~(\ref{S2}) in terms of the coordinates $\theta$ and $\phi$ as follows:
\begin{eqnarray}
     \frac{d\zeta_m}{dt}&=&e^{i\phi_m}\exp\left[{-\int_{0}^{\theta_m}\frac{du}{\alpha-\cos{u}}}\right]\;i\frac{d\phi_m}{dt}+e^{i\phi_m}\exp\left[{-\int_{0}^{\theta_m}\frac{du}{\alpha-\cos{u}}}\right]\;\frac{-1}{\alpha-\cos \theta_m}\;\frac{d \theta_m}{dt} \nonumber\\
     &=&\left[\frac{-1}{\alpha-\cos {\theta_m}}\frac{d\theta_m}{dt}+i\frac{d\phi_m}{dt}\right]\zeta_m. \label{S7}
\end{eqnarray}
Substituting the expression for the conformal factor $\lambda$  of Eq.~(\ref{cnfactor}) of the main text in  Eq.~(\ref{S2}) and comparing with Eq.~(\ref{S7}) we get
\beqa
 \left[\frac{-1}{\alpha-\cos {\theta_m}}\frac{d\theta_m}{dt}+i\frac{d\phi_m}{dt}\right]\zeta_m &=& \frac{2i|\zeta_m|^2}{\left[R-r\cos\theta_m\right]^2}\frac{\partial \psi}{\partial\bar{\zeta}_m}\nn\\
 \Rightarrow \left[\frac{-1}{\alpha-\cos {\theta_m}}\frac{d\theta_m}{dt}+i\frac{d\phi_m}{dt}\right] &=&  \frac{2i}{\left(R-r\cos\theta_m\right)^2}\left[\bar{\zeta}_m\frac{\partial \psi}{\partial\bar{\zeta}_m}\right]\nn
  \\
&=&\frac{2i}{\left(R-r\cos\theta_m\right)^2}\left[\Re\left(\bar{\zeta}_m\frac{\partial \psi}{\partial\bar{\zeta}_m}\right) + i~ \Im\left(\bar{\zeta}_m\frac{\partial \psi} {\partial\bar{\zeta}_m}\right)\right].\label{S8}\\
    \nn
    \eeqa
Comparing real and imaginary parts on both sides of Eq.~(\ref{S8}), we arrive at
    \beqa
    \frac{1}{\alpha-\cos {\theta_m}}\frac{d\theta_m}{dt}= \frac{2}{\left(R-r\cos\theta_m\right)^2} \Im \left(\bar{\zeta}_m\frac{\partial \psi}{\partial\bar{\zeta}_m}\right) \nn\\
    \frac{d\phi_m}{dt}= \frac{2}{\left(R-r\cos\theta_m\right)^2} \Re \left(\bar{\zeta}_m\frac{\partial \psi}{\partial\bar{\zeta}_m}\right).
    \label{S9}
    \eeqa
Now the task is to insert the streamfunction  $\psi$ from Eq.~(\ref{strm}) of main text on the R.H.S. Since $\psi$ is real, we have
   \begin{eqnarray}
       \Re \left(\bar{\zeta}_m\frac{\partial \psi}{\partial\bar{\zeta}_m}\right)&=&\frac{1}{2}\left[\bar{\zeta}_m\frac{\partial \psi}{\partial\bar{\zeta}_m}+{\zeta}_m\frac{\partial \psi}{\partial{\zeta}_m}\right],\quad
    \Im \left(\bar{\zeta}_m\frac{\partial \psi}{\partial\bar{\zeta}_m}\right)=\frac{1}{2i}\left[\bar{\zeta}_m\frac{\partial \psi}{\partial\bar{\zeta}_m}-{\zeta}_m\frac{\partial \psi}{\partial{\zeta}_m}\right]\nn
    \eeqa
     where
    \beqa
     \bar{\zeta}_m\frac{\partial \psi}{\partial\bar{\zeta}_m}&=&
     \left[\sum_{j\neq m}^N \Gamma_j\bar{\zeta}_m \frac{\partial}{\partial\bar{\zeta}_m}G_H(\zeta_m,\zeta_j)+\frac{1}{2}\Gamma_m\bar{\zeta}_m\frac{\partial}{\partial\bar{\zeta}_m}R_m\right]\nn\\
     {\zeta}_m\frac{\partial \psi}{\partial{\zeta}_m}&=&
     \left[\sum_{j\neq m}^N \Gamma_j{\zeta}_m \frac{\partial}{\partial{\zeta}_m}G_H(\zeta_m,\zeta_j)+\frac{1}{2}\Gamma_m{\zeta}_m\frac{\partial}{\partial{\zeta}_m}R_m\right].\label{S10}\\
\nn\end{eqnarray}
 Thus, we need to compute the quantities $\zeta_m\frac{\partial G_H}{\partial \zeta_m},\bar{\zeta}_m\frac{\partial G_H}{\partial \bar{\zeta}_m},\zeta_m\frac{\partial R_m}{\partial \zeta_m}, \bar{\zeta}_m\frac{\partial R_m}{\partial \bar{\zeta}_m}$. Inserting the hydrodynamic Green's function Eq.~(\ref{GH}) and Robin function Eq.~(\ref{robin}) of main text, both of which are real functions, we arrive at (this requires a series of computations which will be presented below)
 \begin{eqnarray}
          \zeta_m\frac{\partial G_{H_{mj}}}{\partial\zeta_m}&=&\frac{K\left(\zeta_m/\zeta_j\right)}{4\pi}+\frac{\alpha\theta_m-\sin\theta_m}{8\pi^2\alpha}+\frac{r_c(\theta_j)}{8\pi^2A}-\frac{1}{8\pi}\nn\\
          \bar{\zeta}_m\frac{\partial G_{H_{mj}}}{\partial\bar{\zeta}_m}&=&\overline{\frac{K\left(\zeta_m/\zeta_j\right)}{4\pi}}+\frac{\alpha\theta_m-\sin\theta_m}{8\pi^2\alpha}+\frac{r_c(\theta_j)}{8\pi^2A}-\frac{1}{8\pi}\nn\\
         \zeta_m\frac{\partial R_m}{\partial \zeta_m}&=&\bar{\zeta}_m\frac{\partial R_m}{\partial \bar{\zeta}_m}= \frac{\alpha\theta_m-\sin\theta_m}{4\pi^2\alpha}+\frac{r_c(\theta_m)}{4\pi^2A}+\frac{\sin\theta}{4\pi}.\label{S11}
\end{eqnarray}
Inserting these expressions back into Eq.~(\ref{S10}), we find that Eq.~(\ref{S9}) takes the form of the vortex dynamical equations Eq.~(\ref{dyneq}) of the main text. All that remains is to systematically compute each of the terms described in Eq.~(\ref{S11}), which is outlined below:\\\\
\underline{Showing~~ $\zeta_m\frac{\partial G_{H_{mj}}}{\partial\zeta_m}=\frac{K\left(\zeta_m/\zeta_j\right)}{4\pi}+\frac{\alpha\theta_m-\sin\theta_m}{8\pi^2\alpha}+\frac{r_c(\theta_j)}{8\pi^2A}-\frac{1}{8\pi}$}.\\
Inserting the Green's function $G_H$ from Eq.~(\ref{GH}) of the main text, we find
\begin{eqnarray}
    \zeta_m\frac{\partial G_H (\zeta_m, \zeta_j)}{\partial \zeta_m}&=&\zeta_m\frac{\partial}{\partial \zeta_m}\left[\frac{1}{2\pi}\log\left|P\left(\frac{\zeta_m}{\zeta_j}\right)\right|\right]+\zeta_m\frac{\partial \varsigma(\eta_m) }{\partial \zeta_m}\nonumber\\
    & &+\zeta_m\frac{\partial }{\partial \zeta_m}\left[\left(\frac{\log|\zeta_m|}{4\pi^2A}-\frac{1}{4\pi}\right)\log|\zeta_m|\right].\label{G4}
 \end{eqnarray}
 We will now evaluate each of the terms arising on the R.H.S of Eq.~(\ref{G4}) separately.\\\\
 \underline{First term $\zeta_m\frac{\partial}{\partial \zeta_m}\left[\frac{1}{2\pi}\log\left|P\left(\frac{\zeta_m}{\zeta_j}\right)\right|\right]$}\\\\
Introducing the function $K(\zeta_m)$  which is the logarithmic derivative of the Schottky-Klein prime function $P(\zeta_m)$ defined as
\begin{equation}
 K(\zeta_m)=\zeta_m\frac{\partial }{\partial \zeta_m}\log P(\zeta_m)=\frac{\zeta_m}{P(\zeta_m)}\frac{\partial }{\partial\zeta_m}P(\zeta_m),\label{G5} 
\end{equation}
we can express the first term of Eq.~(\ref{G4}) as
\begin{eqnarray}
    \zeta_m\frac{\partial}{\partial \zeta_m}\left[\frac{1}{2\pi}\log\left|P\left(\frac{\zeta_m}{\zeta_j}\right)\right|\right]&=&\frac{1}{4\pi}K\left(\frac{\zeta_m}{\zeta_j}\right).\label{G6}
\end{eqnarray}
\underline{Second Term ~$\zeta_m\frac{\partial \varsigma(\eta_m) }{\partial \zeta_m}$}\\ 
The function $\varsigma(\eta)$ is defined in Eq.~(\ref{GH}) of main text, which we reproduce here
\begin{equation}
    \varsigma(\eta)=\frac{A}{2\pi^2} \text{Re}\left[Li_2(c^{-1}\eta)\right]-\frac{1}{2\pi^2\alpha}\log|\eta-c|+\frac{A}{8\pi^2}(\log\eta)^2.\label{function}
\end{equation}
\begin{comment}
Lets define the quantity 
\begin{equation}
    \zeta\frac{\partial}{\partial\zeta}\varsigma(\eta)=f(\eta)
\end{equation}
where 
\begin{equation}
    f(\eta)=\frac{-i}{8\pi^2}\left[\log\left[\frac{\eta-c}{\eta-c^{-1}}\right]+\frac{1}{\alpha A}\left[\frac{c}{\eta-c}+\frac{c^{-1}}{\eta-c^{-1}}\right]+\frac{i}{8\pi^2}\left[\log(-c)\right]-\frac{1}{\alpha A}\right].\label{function f}
\end{equation}
\end{comment}
It will be helpful to represent the first and second terms of Eq.~(\ref{function}) via integral representations as follows:
\begin{equation}
   \Re \left[Li_2(c^{-1}\eta)\right]=-\frac{1}{2}\int \log\left[\frac{\eta-c}{\eta-c^{-1}}\right]\frac{d\eta}{\eta}-\frac{1}{4}(\log\eta)^2+\frac{1}{2}\,\log\eta\, \log(-c)
\label{term A}\end{equation}
and
\begin{equation}
\log|\eta-c|=\frac{1}{2}\int\left[\frac{c}{\eta-c}+\frac{c^{-1}}{\eta-c^{-1}}\right]\frac{d\eta}{\eta}+\frac{1}{2}\log(-c)+\frac{1}{2}\log\eta.
\label{term B}\end{equation}
Inserting Eq.~(\ref{term A}) and Eq.~(\ref{term B}) back to Eq.~(\ref{function}) we get,
\begin{eqnarray}
    \varsigma(\eta)&=&-\frac{A}{4\pi^2}\int\left[\log\left[\frac{\eta-c}{\eta-c^{-1}}\right]+\frac{1}{\alpha A}\left[\frac{c}{\eta-c}+\frac{c^{-1}}{\eta-c^{-1}}\right]\right]\frac{d\eta}{\eta}\nn\\
    &&+\frac{A}{4\pi^2}\left[\log(-c)-\frac{1}{\alpha A}\right]\log \eta-\frac{1}{4\pi^2\alpha}\log(-c).\nn\\
\nn\end{eqnarray}
We can thus write the function $\varsigma(\eta)$ in terms of an integral representation 
\begin{equation}
    \varsigma(\eta)=-2iA\int \frac{f(\eta)}{\eta}d\eta-\frac{1}{4\pi^2\alpha}\log(-c).
\label{modified function}\end{equation}
where 
\begin{equation}
    f(\eta)=\frac{-i}{8\pi^2}\left[\log\left[\frac{\eta-c}{\eta-c^{-1}}\right]+\frac{1}{\alpha A}\left[\frac{c}{\eta-c}+\frac{c^{-1}}{\eta-c^{-1}}\right]+\frac{i}{8\pi^2}\left[\log(-c)\right]-\frac{1}{\alpha A}\right].\label{function f}
\end{equation}
Now, we compute the derivative $\zeta_m\frac{\partial}{\partial\zeta_m}\varsigma(\eta_m)$ as follows:
\begin{equation}
    \zeta_m\frac{\partial}{\partial\zeta_m}\varsigma(\eta_m)=\zeta_m\frac{\partial\eta_m}{\partial\zeta_m}\frac{\partial\varsigma(\eta_m)}{\partial\eta_m}= \zeta_m \left(\frac{i\eta_m}{2A\zeta_m}\right) \frac{\partial\varsigma(\eta_m)}{\partial\eta_m}= \left(\frac{i\eta_m}{2A}\right) \frac{\partial\varsigma(\eta_m)}{\partial\eta_m} =f(\eta_m)
\label{modified derivative}\end{equation}
where in the last step we differentiated Eq.~(\ref{modified function}) w.r.t  $\eta$. All that remains is to express $f(\eta)$ in terms of $\theta$, for which we proceed as follows
\begin{equation}
    \frac{df}{d\theta_m}=\frac{df}{d\eta_m}\frac{d\eta_m}{d\theta_m} =\left( \frac{iA}{8\pi^2\alpha r^2\eta_m}(R-r\cos\theta_m)^2\right) \left(\frac{-i\eta_m}{A\left[\alpha-\cos\theta_m\right]}\right)= \frac{\alpha-\cos\theta_m}{8\pi^2\alpha}=\frac{d}{d\theta_m}\left[\frac{\alpha\theta_m-\sin\theta_m}{8\pi^2\alpha}\right]\nn
    \end{equation}
    where in the last few steps we used an identity relating $\eta$ and $\theta$ given in Appendix \ref{app4}.
    It thus follows that
    \beqa
    \zeta_m\frac{\partial}{\partial\zeta_m}\varsigma(\eta_m)&=&f\left(\eta_m[\theta_m]\right)=\frac{\alpha\theta_m-\sin\theta_m}{8\pi^2\alpha}.\label{G7}
\eeqa
\underline{Third Term~$\zeta_m\frac{\partial }{\partial \zeta_m}\left[\left(\frac{\log|\zeta_m|}{4\pi^2A}-\frac{1}{4\pi}\right)\log|\zeta_m|\right]$} \\
This is easy to evaluate and leads to
  \beqa
   \zeta_m\frac{\partial }{\partial \zeta_m}\left[\left(\frac{\log|\zeta_j|}{4\pi^2A}-\frac{1}{4\pi}\right)\log|\zeta_m|\right]&=&\frac{r_c(\theta_j)}{8\pi^2A}-\frac{1}{8\pi}.\label{G8}\\
\nn\end{eqnarray}
Combining Eqs.~(\ref{G6}, \ref{G7}, \ref{G8}) and substituting in Eq.~(\ref{G4}) we finally have the required result ie.  
\begin{eqnarray}
    \zeta_m\frac{\partial G_H}{\partial \zeta_m}=\frac{1}{4\pi}K\left(\frac{\zeta_m}{\zeta_j}\right)+\frac{\alpha\theta_m-\sin\theta_m}{8\pi^2\alpha} +\frac{r_c(\theta_j)}{8\pi^2A}-\frac{1}{8\pi}.
    \label{G9}
\end{eqnarray}
Taking complex conjugate of this equation and using the reality of $G_H$, we arrive at the second equation of Eq.~(\ref{S11}).\\\\
\underline{Showing $\zeta_m\frac{\partial R_m}{\partial \zeta_m}=\bar{\zeta}_m\frac{\partial R_m}{\partial \bar{\zeta}_m}= \frac{\alpha\theta_m-\sin\theta_m}{4\pi^2\alpha}+\frac{\log|\zeta_m|}{4\pi^2A}-\frac{\sin\theta_m}{4\pi}$}\\
We proceed by inserting the expression for the Robin function from the main text, Eq.~(\ref{robin}), 
\begin{eqnarray}
   \zeta_m\frac{\partial R_m}{\partial \zeta_m}&=\zeta_m\frac{\partial \varsigma(\eta_m) }{\partial \zeta_m}+\zeta_m\frac{\partial }{\partial \zeta_m}\left[\frac{(\log|\zeta_m|)^2}{4\pi^2A}-\frac{\log|\zeta_m|}{4\pi}\right]-\zeta_m\frac{\partial }{\partial \zeta_m}\left( \int_{0}^{\theta_m}\frac{du}{4\pi^2\alpha}\frac{\alpha(u+\pi)-\sin u}{\alpha-\cos u}\right)\nn\\&-\frac{\zeta_m}{2\pi}\frac{\partial \log\left[\lambda(\zeta_m)|\zeta_m|\right] }{\partial \zeta_m}
   \label{R0}
\end{eqnarray}
We evaluate each term in Eq.~(\ref{R0}) separately.\\ 
\underline{first term $\zeta_m\frac{\partial \varsigma(\eta_m) }{\partial \zeta_m}$}\\
We have already shown in Eq.~(\ref{G7}) that
\begin{equation}
    \zeta_m\frac{\partial \varsigma(\eta_m) }{\partial \zeta_m}=\frac{\alpha\theta_m-\sin\theta_m}{8\pi^2\alpha}.\label{R1}\\
\end{equation}
\underline{second term $\zeta_m\frac{\partial }{\partial \zeta_m}\left[\frac{(\log|\zeta_m|)^2}{4\pi^2A}-\frac{\log|\zeta_m|}{4\pi}\right]$}\\
We have already seen from~(\ref{G8}) that
\begin{eqnarray}
    \zeta_m\frac{\partial }{\partial \zeta_m}\left[\frac{\log|\zeta_m|^2}{4\pi^2A}-\frac{1}{4\pi}\log|\zeta_m|\right]=\frac{\log|\zeta_m|}{4\pi^2A}-\frac{1}{8\pi}.\label{R2}
   \end{eqnarray}
\underline{Third term $\zeta_m\frac{\partial }{\partial \zeta_m}\left( \int_{0}^{\theta_m}\frac{du}{4\pi^2\alpha}\frac{\alpha(u+\pi)-\sin u}{\alpha-\cos u}\right)$}\\
We use chain rule
\beqa
    &\zeta_m\frac{\partial}{\partial \zeta_m}\left( \int_{0}^{\theta_m}\frac{du}{4\pi^2\alpha}\frac{\alpha(u+\pi)-\sin u}{\alpha-\cos u}\right)=\zeta_m\frac{\partial \eta_m}{\partial \zeta_m}\frac{\partial \theta_m}{\partial \eta_m}\frac{\partial }{\partial \theta_m} \left( \int_{0}^{\theta_m}\frac{du}{4\pi^2\alpha}\frac{\alpha(u+\pi)-\sin u}{\alpha-\cos u}\right)\nn\\
    &=\zeta_m \left(\frac{i\eta_m}{2A\zeta_m} \right) \left(\frac{-A}{i\eta_m}\left[\alpha-\cos\theta_m\right]\right) \left(\frac{1}{4\pi^2\alpha}\left[\frac{\alpha(\theta_m+\pi)-\sin\theta_m}{\alpha-\cos\theta_m}\right]\right)\nn\\
    &=-\left(\frac{1}{8\pi}+\frac{\alpha\theta_m-\sin\theta_m}{8\pi^2\alpha}\right)
    \label{R3}
\eeqa
\underline{Fourth term $\frac{\zeta_m}{2\pi}\frac{\partial \log\left[\lambda(\zeta_m)|\zeta_m|\right] }{\partial \zeta_m}$}\\
We again use chain rule to write
\beqa
\frac{1}{2\pi} \zeta_m\frac{\partial }{\partial \zeta_m}\log\left[\lambda(\zeta_m)|\zeta_m|\right]&=& \frac{1}{2\pi} \zeta_m\frac{\partial \eta_m }{\partial \zeta_m}\frac{\partial}{\partial \eta_m}\log\left[\lambda(\zeta_m)|\zeta_m|\right].\label{fourth term}\\
\nn\eeqa
To proceed further, we have to express the conformal factor $\lambda$ appearing on the R.H.S  in terms of $\eta_m$ , where $\eta_m=|\zeta_m|^{i/A}$. This is easily achieved by using the identity discussed in Appendix \ref{app4}, we find
\begin{equation}
    \lambda= \frac{2r}{A^2|\zeta|}\frac{\eta}{(\eta-c)(\eta-c^{-1})}.\label{R8}
\end{equation}    
Plugging Eq.~(\ref{R8}), Eq.~(\ref{fourth term}) takes the form, 
\beqa
&\frac{1}{2\pi} \zeta_m\frac{\partial }{\partial \zeta_m}\log\left[\lambda(\zeta_m)|\zeta_m|\right] =\frac{1}{2\pi}\zeta_m\frac{\partial \eta_m }{\partial \zeta_m}\frac{\partial}{\partial \eta_m}\log \left[\frac{2r}{A^2}\frac{\eta}{(\eta-c)(\eta-c^{-1})}\right]\nn\\
&= \frac{1}{2\pi} \frac{i}{2A}\left[1-\frac{\eta_m}{\eta_m-c}-\frac{\eta_m}{\eta_m-c^{-1}}\right]
\nn\\
&=-\frac{\sin \theta_m}{4\pi}
\label{R4}
\eeqa
where in the last step we used Eq.~(\ref{rcdef})  and Appendix \ref{app4} to express $\eta_m$ in terms of $\theta_m$. Collecting Eqs.~(\ref{R1}), (\ref{R2}), (\ref{R3}) and (\ref{R4}) we get the required result
\begin{eqnarray}
 \zeta_m\frac{\partial R_m}{\partial \zeta_m}&=& \left(\frac{\alpha\theta_m-\sin\theta_m}{8\pi^2\alpha}\right)+ \left(\frac{\log|\zeta_m|}{4\pi^2A}-\frac{1}{8\pi} \right)+ \left(\frac{1}{8\pi}+\frac{\alpha\theta_m-\sin\theta_m}{8\pi^2\alpha}\right) + \frac{\sin\theta_m}{4\pi}\nn\\
 &=&\frac{\alpha\theta_m-\sin\theta_m}{4\pi^2\alpha}+\frac{\log|\zeta_m|}{4\pi^2A}+\frac{\sin\theta_m}{4\pi}.\nn
\nn\end{eqnarray}
It also follows that $\zeta_m\frac{\partial R_m}{\partial \zeta_m}=\bar{\zeta}_m\frac{\partial R_m}{\partial \bar{\zeta}_m}$.
\section{Relation between $\eta$ and $\theta$}
\label{app4}
In this section, we will establish a useful identity 
\beqa
   \frac{2r\eta/A^2}{\eta^2+2\alpha \eta+1}=\frac{\frac{2r\eta}{{A^2}}}{(\eta-c)(\eta-c^{-1})}=R-r\cos\theta
\eeqa
relating the variable $\eta$ and  the corodinate $\theta$  of the torus, which has been used in Appendix \ref{app3}. The first equality follows immediately from the definition of $c$ in main text Eq.~(\ref{param}). We show the third equality below. Starting from the LHS of the above identity, 
\begin{eqnarray}
\frac{2r\eta/A^2}{\eta^2+2\alpha\eta+1}&=& \frac{2r}{A^2}\left[\frac{1}{\eta+2\alpha+\eta^{-1}}\right] \nonumber \\
&=& \frac{r}{A^2}\left[\frac{1}{\alpha+\cos\left(\frac{r_c\left(
{\theta}\right)}{A}\right)}\right] \label{d1}
\end{eqnarray}
where $r_c(\theta)$ is defined in Eq.~(\ref{rcdef}) of main text and reproduced again for convenience,
\begin{eqnarray}
   r_c(\theta) =-2 A\arctan\left(A (1+\alpha)\tan \frac{\theta}{2}\right)\nn
\end{eqnarray}
Plugging the function $r_c(\theta)$ in  Eq.~(\ref{d1}) we obtain the desired relation between $\eta$ and $\theta$ after straightforward trigonometric manipulations.


\begin{thebibliography}{99}
%moffattarefsaffman
\bibitem{moffatt} Keith Moffatt, \textit{Vortex Dynamics: the legacy of Kelvin and Helmholtz}, IUTAM Symposium on Hamiltonian Dynamics, Vortex Structures, Turbulence. IUTAM Bookseries, 6, Springer (2008)

\bibitem{aref} H. Aref, \textit{Point vortex dynamics: A classical mathematics playground}, Journal of Mathematical Physics, 48, 6, 065401 (2007)

\bibitem{aref1} H. Aref, \textit{Integrable, Chaotic, and Turbulent Vortex
Motion in Two-Dimensional Flows}, Ann. Rev. Fluid Mech. 15, 345 (1983)

 \bibitem{saffman} P.G. Saffman, \textit{Vortex dynamics}. Cambridge University Press (1993)

 \bibitem{lin1} C.C. Lin, \textit{ On the Motion of Vortices in Two Dimensions: I. Existence of the Kirchhoff-Routh Function}, Proc. Natl. Acad. Sci. U S A. 27, 570-575 (1941)

\bibitem{lin2} C.C. Lin, \textit{ On the Motion of Vortices in Two Dimensions: II. Some Further Investigations on the Kirchhoff-Routh Function}, Proc. Natl. Acad. Sci. U S A, 27, 575-577 (1941)

 
\bibitem{bg} V.A. Bogomolov, \textit{ Dynamics of vorticity at a sphere, }Fluid Dyn 12, 863–870 (1977)

\bibitem{hally} D. Hally, \textit{Stability of streets of vortices on surfaces of revolution with a reflection symmetry},J. Math. Phys. 21, 211-217 (1980)
    
\bibitem{kimok} Y. Kimura and H. Okamoto, \textit{Vortex Motion on a Sphere}, Journal of the Physical Society of Japan, 56, 4203-4206, 1987.
    
\bibitem{kimura} Y. Kimura, \textit{Vortex motion on surfaces with constant curvature}, Proc. R. Soc. Lond, A.455:245–259 (1999)

\bibitem{khesin2024} T.D. Drivas, D. Glukhovskiy and B.Khesin, \textit{Singular Vortex Pairs Follow Magnetic Geodesics}, International Mathematics Research Notices, Volume 2024, Issue 14, Pages 10880–10894 (2024)



\bibitem{newton} P.K. Newton, \textit{ The N-Vortex Problem : Analytical Techniques,} Springer New York (2001) 

\bibitem{crowdymarshall} D.G. Crowdy and J. Marshall, \textit{ Analytical Formulae for the Kirchhoff-Routh Path Function in Multiply Connected Domains}, Proceedings: Mathematical, Physical and Engineering Sciences, 461(2060), 2477–2501 (2005)


%turnervoigt
\bibitem{turner} A.M. Turner, V. Vitelli and D.R. Nelson, \textit{Vortices on curved surfaces}, Rev. Mod. Phys. 82, 1301-1348 (2010)

    
\bibitem{voigt} R. Reuther and A. Voigt, \textit{The interplay of curvature and vortices in flow on curved surfaces, SIAM J.Multiscale Model.Simul.} 13, 632-643 ( 2014)
     



%boatto  
\bibitem{boattok} S.Boatto and J.Koiller, \textit{Vortex on closed surfaces,} Fields Institute Communications 73, 185–237 (2015)

\bibitem{boattod} D.G.Dritschel and S.Boatto, \textit{The motion of point vortices on closed surfaces}, Proc. R. Soc. A.47120140890 (2015)

\bibitem{moffat2014} H.K. Moffat, \textit{Helicity and singular structures in fluid}, PNAS, Volume 111, Issue 10, March 2014, Pages 3663-3670 (2014)
\bibitem{ershkov2016} S.V. Ershkov, \textit{About existence of stationary points for the Arnold-Beltrami-Childress (ABC) flow}, Applied Mathematics and Computation, vol. 276, pp. 379-383 (2016)
 %greenmarshall

\bibitem{grms} CC. Green and JS. Marshall  \textit{ Green’s function for the Laplace-Beltrami operator on a toroidal
surface}, Proc. R. Soc. A 469, 20120479 (2012)  


%sakajo
\bibitem{sakajo2009} T. Sakajo, \textit{Equation of motion for point vortices in multiply connected circular domains }, Proc. R. Soc. A.4652589–2611, 2009.  

 \bibitem{newtonsakajo1} P.K. Newton and T. Sakajo, \textit{The N-vortex problem on a rotating sphere. III. Ring configurations coupled to a background field}, Proc. R. Soc. A.463:961–977, 2007.

 \bibitem{newtonsakajo2} P.K. Newton and T. Sakajo, \textit{Point vortex equilibria and optimal packings of circles on a sphere}, Proc. R. Soc. A.467, 1468–1490 (2011).
 
\bibitem{nelsonsakajo} R. Nelson and T. Sakajo, \textit{ Trapped vortices in multiply connected domains}, Fluid Dynamics Research. 46, 061402 (2014). 

\bibitem{sakajo2016} T. Sakajo and Y. Shimizu, \textit{Point vortex interactions on a toroidal surface}, Proc. R. Soc. A 472: 20160271 (2016)

\bibitem{sakajo2018} T. Sakajo and Y. Shimizu, \textit{Toroidal Geometry Stabilizing a Latitudinal Ring of Point Vortices on a Torus}, J Nonlinear Sci, 28:1043–1077 (2018)

\bibitem{sakajo2019} T. Sakajo, \textit{Vortex crystals on the surface
of a torus}, Phil. Trans. R.
Soc. A 377: 20180344 (2019)

\bibitem{sakajo2023} V. Krishnamurthy and T. Sakajo, \textit{ The N -vortex problem in a doubly periodic rectangular domain with constant background vorticity}, Physica D: Nonlinear Phenomena.448.133728 (2023) 



 \bibitem{sakajo2025} T. Sakajo and Z. Changjun, \textit{Steady vortex patches on flat torus with a constant background vorticity}, arXiv:2501.04271.
    



%rsworks + Lauga
\bibitem {rs2021}  R. Samanta and N. Oppenheimer, \textit{Vortex Flows and Streamline Topology in Curved Biological Membranes}, Physics of Fluids, 33, 5, 051906 (2021)
\bibitem{rs2022} S. Bagaria and R. Samanta, \textit{Dynamics of  force dipoles in curved fluid membranes}, Phys. Rev. Fluids 7, 093101 (2022)
\bibitem{rs2023} S. Jain and R. Samanta, \textit {Force dipole interactions in tubular fluid membranes}, Physics of Fluids 35, 071901 (2023)

\bibitem {rs2025} U. Maurya, S. Gavva, A. Saha and R. Samanta, \textit{Vortex Dynamics in Tubular Fluid Membranes}, Physics of Fluids, 37, 7 (2025) 

\bibitem{lauga2025} M. Vona and E. Lauga, \textit{Rotational mobility in spherical membranes: the interplay between Saffman–Delbrück length and inclusion size}, Proc. R. Soc. A.481, 20240473 (2025)
%yeolushi

\bibitem{lushi} E.Lushi and P.M.Vlahovska, \textit{Periodic and Chaotic Orbits of Plane-Confined Micro-rotors in Creeping Flows}, J Nonlinear Sci 25, 1111–1123, 2015.
    
\bibitem{yeo}  K.Yeo, E.Lushi and P.M.Vlahovska, \textit{Collective dynamics in a binary mixture of hydrodynamically coupled microrotors}, Phys. Rev. Lett, 114(18), Article 188301 (2015)

\bibitem{sh1} N. Oppenheimer, D.B. Stein, and M.J. Shelley. \textit{Rotating membrane inclusions
crystallize through hydrodynamic and steric interactions}, Phys. Rev. Lett., 123:148101 (2019)

\bibitem{sh2} N. Oppenheimer, D.B. Stein, M.Y.B. Zion, and M.J. Shelley, \textit{ Hyperuniformity and phase enrichment in vortex and rotor assemblies}, Nature Communications,
13,1, (2022).
  
%giomi et al
\bibitem{giomi2019} D.J.G. Pearce, P.W. Ellis, A. Fernandez-Nieves and L. Giomi, \textit{Geometrical Control of Active Turbulence in Curved Topographies}, Physical Review Letters, 122, 168002 (2019)
\bibitem{giomi2008}  L. Giomi and M.J. Bowick, \textit{Elastic theory of defects in toroidal crystals}, Eur. Phys. J. E 27, 275–296 (2008)
\bibitem{giomi2008pre} L. Giomi and M.J. Bowick, \textit{Defective ground states of toroidal crystals}, Physical Review E 78, 010601 (2008)
\bibitem{giomi2024} J. Rojo-González, L.N. Carenza, A.D.L. Cotte, L.A. Hoffmann, L.Giomi, A. Fernandez-Nieves, \textit{Defect-populated configurations in nematic solid tori and cylinders}, Physical Review Research 6, L012065 (2024)

 %superfluid 
 \bibitem{feynman} R. P. Feynman,\textit{ Application of quantum mechanics to liquid Helium}, Progress in low
Temperature Physics (North Holland, Amsterdam) 117-53 (1955)

\bibitem{emn} A. Corrada-Emmanuel, \textit{Exact solution for superfluid film vortices on a torus}, Phys. Rev.
Lett. 72, 681–684 (1994)

\bibitem{machta} J Machta and R.A. Guyer, \textit{Superfluid films in porous media}, Phys. Rev. Lett. 60, 2054 (1988)

\bibitem{fetter} N.E. Guenther, P. Massignan, A.L. Fetter, \textit{ Superfluid vortex dynamics on a torus and other toroidal surfaces of revolution},	Phys. Rev. A 101, 053606 (2020)


\bibitem{abanov} P. Weigmann and A. Abanov, \textit{Anomalous Hydrodynamics of Two-Dimensional Vortex Fluids}, PRL 113, 034501 (2014)
\bibitem{vsc} Gauthier et al., \textit{Giant vortex clusters in a
two-dimensional quantum fluid}, Science 364, 1264–1267 (2019) 


%torus experiments
\bibitem{trap1} A. Chakraborty, S. R.Mishra, S. P. Ram, S. K. Tiwari, and H. S
Rawat, \textit{A toroidal trap for cold
Rb atoms using an rf-dressed
quadrupole trap}, J. Phys. B 49, 075304 (2016).
\bibitem{trap2} A. L. Gaunt, T. F. Schmidutz, I. Gotlibovych, R. P. Smith,
and Z. Hadzibabic, \textit{Bose-Einstein Condensation of Atoms in a uniform potential}, Phys. Rev. Lett. 110, 200406 (2013)
%Andrea tononi
\bibitem{andrea1} A. Tononi and L. Salasnich, \textit{Low-dimensional quantum gases in curved geometries}, Nat Rev Phys 5, 398–406 (2023).

\bibitem{andrea2} A. Tononi and G. E Astrakharchik, \textit{Bose-Einstein condensation on axially symmetric surfaces}, Phys. Rev. A 112, 023302 (2025).

\bibitem{andrea3} A. Tononi, L. Salasnich and A. Yakimenko, \textit{Quantum vortices in curved geometries}, AVS Quantum Sci. 6, 030502 (2024).
\bibitem{iss1} N. Lundblad, R. A. Carollo, C. Lannert, M. J. Gold, X. Jiang,
D. Paseltiner, N. Sergay, and D. C. Aveline, \textit{Shell potentials for
microgravity Bose-Einstein condensates}, NPJ Microgravity 5, 30
(2019)

\bibitem{iss2} G. Condon, M. Rabault, B. Barrett, L. Chichet, R. Arguel,
H. Eneriz-Imaz, D. Naik, A. Bertoldi, B. Battelier, P. Bouyer,
and A. Landragin, \textit{All-Optical Bose-Einstein Condensates in
Microgravity}, Phys. Rev. Lett. 123, 240402 (2019)


%crowdyskprime
\bibitem{crowdyskpaper} D.G. Crowdy, E.H. Kropf, C.C. Green and M.M.S. Nasser, \textit{The Schottky–Klein prime function: a theoretical and computational tool for applications}, IMA Journal of Applied Mathematics  81, 589–628 (2016)

\bibitem{crowdybook} D.G. Crowdy, \textit{Solving problems in multiply connected domains}, CBMS-NSF Regional Conference Series in Applied Mathematics (2020)

%harmonic field addition
\bibitem{h1} T. D Drivas and T. M Elgindi, \textit{Singularity formation in the incompressible Euler equation in finite and infinite time}, EMS Surveys in Mathematical Sciences 10, no. 1, 1–100 (2023)
\bibitem{h2} H. Yin, M. Nabizadeh, B. Wu, S. Wang, A. Chern, \textit{Fluid cohomology}, ACM Trans. Graph., Vol. 42, No. 4, Article 126. Publication date: August 2023

\bibitem{h3} C. Grotta-Ragazzo, B. Gustafsson, and  J. Koiller, \textit{On the Interplay Between Vortices and Harmonic Flows: Hodge Decomposition of Euler’s Equations in 2d}, Regul. Chaot. Dyn. 29, 241–303 (2024)


%math

\bibitem{akh} NI Akhiezer, \textit{Elements of the theory of elliptic functions}, Providence, RI: American
Mathematical Society (1990)


\bibitem{qg0} F.H. Jackson, \textit{The Basic Gamma-Function and the Elliptic Functions}, Proceedings of the Royal Society of London. Series A, 76 (508), The Royal Society: 127–144 (1905)

\bibitem{qg1} G.E. Andrews, \textit{ W. Gosper's Proof that $\lim _{q \rightarrow 1^{-}} \Gamma_q(x)=\Gamma(x)$}, Appendix A in qSeries: Their Development and Application in Analysis, Number Theory, Combinatorics, Physics, and Computer Algebra, Providence, RI: Amer. Math. Soc., p. 11 and 109 (1986).

\bibitem{qg2}
G. Gasper and M. Rahman, \textit{Basic Hypergeometric Series}, Cambridge University Press (1990)

\bibitem{qg3} R. Koekoek and R.F. Swarttouw, \textit{ The q-Gamma Function and the q-Binomial Coefficient}, in The Askey-Scheme of Hypergeometric Orthogonal Polynomials and its q-Analogue. Delft, Netherlands: Technische Universiteit Delft, Faculty of Technical Mathematics and Informatics Report 98-17, pp. 10-11, (1998).

\bibitem{qg4} W. Koepf, \textit{Hypergeometric Summation: An Algorithmic Approach to Summation and Special Function Identities}, Braunschweig, Germany: Vieweg (1998)

\bibitem{qg5} C. Wenchang, \textit{ Problem 10226 and Solution. A q-Trigonometric Identity }, Amer. Math. Monthly 103, 175-177, (1996).

\bibitem{tokieda} J. Montaldi, A. Souliere and T. Tokeida, \textit{Vortex Dynamics on a Cylinder}, Siam Journal of Applied Dynamical Systems, Vol. 2, No. 3, pp. 417–430 (2003).




\end{thebibliography}
\end{document}